\documentclass[prl,aps,twocolumn,superscriptaddress]{revtex4-1}
\usepackage{amsmath}
\usepackage{color}
\usepackage{bbm}
\usepackage{amssymb}
\usepackage{epsfig}
\usepackage{multirow}
\usepackage{amsbsy}
\usepackage{array}
\usepackage{diagbox}
\usepackage{bm}
\usepackage{extarrows}
\usepackage{graphicx}
\usepackage{appendix}
\usepackage{txfonts}

\usepackage[colorlinks=true,linkcolor=blue,citecolor=blue,urlcolor=blue,bookmarks=false]{hyperref}

\begin{document}

\title{Density matrix renormalization group boosted by Gutzwiller projected wave functions}

\author{Hui-Ke Jin}
\affiliation {Beijing National Laboratory for Condensed Matter Physics $\&$ Institute of Physics, Chinese Academy of Sciences, Beijing 100190, China}

\author{Hong-Hao Tu}
\email{hong-hao.tu@tu-dresden.de}
\affiliation{Institut f\"ur Theoretische Physik, Technische Universit\"at Dresden, 01062 Dresden, Germany}

\author{Yi Zhou}
\email{yizhou@iphy.ac.cn}
\affiliation {Beijing National Laboratory for Condensed Matter Physics $\&$ Institute of Physics, Chinese Academy of Sciences, Beijing 100190, China}
\affiliation{Songshan Lake Materials Laboratory, Dongguan, Guangdong 523808, China}
\affiliation{Kavli Institute for Theoretical Sciences $\&$ CAS Center for Excellence in Topological Quantum Computation, University of Chinese Academy of Sciences, Beijing 100190, China}

\date{\today}

\begin{abstract}
		We propose to boost the performance of the density matrix renormalization group (DMRG) in two dimensions by using Gutzwiller projected states as the initialization ansatz. When the Gutzwiller projected state is properly chosen, the notorious ``local minimum'' issue in DMRG can be circumvented and the precision of DMRG can be improved by orders of magnitude without extra computational cost. Moreover, this method allows to quantify the closeness of the initial Gutzwiller projected state and the final converged state after DMRG sweeps, thereby sheds light on whether the Gutzwiller ansatz captures the essential entanglement features of the actual ground state for a given Hamiltonian. The Kitaev honeycomb model has been exploited to demonstrate and benchmark this new method.
\end{abstract}

\maketitle

{\em Introduction.---} Since its invention by White in 1992~\cite{white1992,white1993}, the density matrix renormalization group (DMRG) has been recognized as the most powerful computational method for studying strongly correlated quantum systems in one dimension~\cite{RMPSchollwock,AdvPhyHallberg,AnnalPhysSchollwock}. Soon after that, it was realized that DMRG can be formulated as a variational method operating within the family of matrix product states (MPSs)~\cite{DMRG2MPS_1,DMRG2MPS_2}. This discovery leads to a deeper and coherent understanding of the inner structure of the DMRG method, as well as its potential and limitations~\cite{verstraete2006,hastings2007}. For instance, it becomes clear that DMRG is only moderately successful when applied to two-dimensional (2D) quantum systems~\cite{stoudenmire2012}: while relatively small systems can be computed with high accuracy, the computational resources required grow exponentially with the system size, making large systems intractable. The sharply distinct performance of DMRG in one and two dimensions originates from the different entanglement scaling in many-body ground states with respect to spatial dimensionality, dictated by the so-called area law~\cite{entangle2D1,entangle2D2,RMPAreaLaws}.

For 2D quantum systems, the common practice of DMRG is to consider lattices with cylindrical boundary conditions and gradually increase the circumference of the cylinder~\cite{stoudenmire2012}. However, the convergence of DMRG to the ground state is not guaranteed due to the presence of local minima in the energy landscape. As a result, the efficiency and accuracy of DMRG highly depend on how initial states are chosen. It is expected that the performance of DMRG can be improved by using  some initial states that capture the essential physics. Actually, Gutzwiller projected wave functions have long been used as variational ansatz for strongly correlated electrons and quantum spin systems, which have proven success in a number of important instances~\cite{Gros89,anderson04,rmp06,QSLRMP}. This raises a very natural question: can one utilize Gutzwiller projected wave functions to improve the performance of DMRG?

Very recently, it was proposed by us~\cite{MPOMPS2} and coworkers~\cite{MPOMPS1} that a Gutzwiller projected state can be efficiently represented as a tensor network and subsequently compressed as an MPS by using the so-called matrix product operator-matrix product state (MPO-MPS) method. This completes the building block of initializing DMRG with Gutzwiller projected states. The accuracy of the MPO-MPS method has already been carefully examined for various one-dimensional systems~\cite{MPOMPS2,MPOMPS1}. Along this line, the present work focuses on (i) sorting out the subtleties of the MPO-MPS method for 2D systems with cylindrical boundary conditions and (ii) analyzing the performance of DMRG initialized with Gutzwiller projected states.

The Kitaev honeycomb model~\cite{Kitaev06}, being a rare exactly solvable example in two dimensions, is used for illustrating our method. Our extensive analysis shows that the MPO-MPS method, with several subtleties taken into account, converts Gutzwiller projected states into MPSs with satisfactory precision and the performance of DMRG is dramatically improved when initialized with these MPSs. We also address a controversial issue on the Kitaev honeycomb model with antiferromagnetic Kitaev interactions and a magnetic field in [111] direction. With a nonzero field, this model is no longer exactly solvable and was claimed to support a disordered state at intermediate field strength~\cite{zhu2018,gohlke2018,hickey2019}. We use our method to analyze early proposed candidate wave functions~\cite{YMLu2018,QHWang2019} and found that although some of them describe actual ground states well in both small and large field limits, all of them seem to fail in the region with intermediate field strength, thus calling for further investigations on the nature of the field-induced disordered state.

\begin{figure}[tb]
	\includegraphics[width=\linewidth]{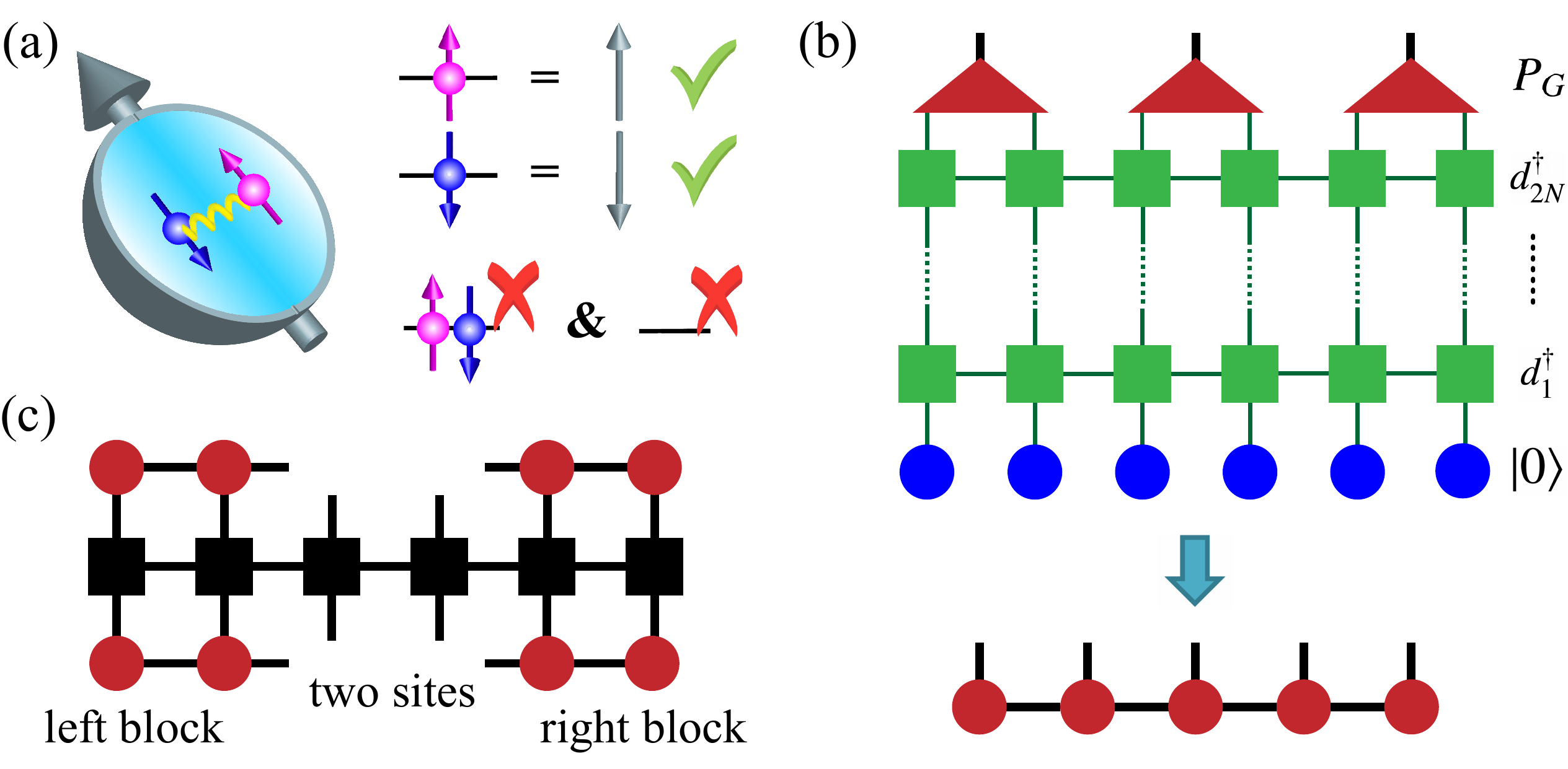}
	\caption{(a) Sketch of parton construction for a quantum spin-1/2 system, where Gutzwiller projection keeps two single-occupied states and removes other components locally.  (b) Convert a Gutzwiller projected wave function into an MPS by the MPO-MPS method. (c) The MPS prepared in (b) serves as an initial state for two-site DMRG.}
	\label{fig:intro}
\end{figure}

{\em Method.---} Throughout this work, we consider spin-1/2 lattice systems and Gutzwiller projected states with singly occupied fermionic partons at each site [see Fig.~\ref{fig:intro}(a)], whereas generalizations to other systems and/or different parton descriptions are straightforward. Our method consists of three main steps:

(1) Construct the Gutzwiller projected state as $|\Psi_G\rangle = P_{G}|\Psi_0\rangle$, where $|\Psi_0\rangle$ is the ground (or excited) state of a quadratic Hamiltonian for fermionic partons and $P_{G}$ is the Gutzwiller projector imposing the single-occupancy constraint.

(2) Convert $|\Psi_G\rangle$ into an MPS by using the MPO-MPS method~\cite{MPOMPS2,MPOMPS1} as illustrated in Fig.~\ref{fig:intro}(b), and keep the bond dimension of the resulting MPS up to $\tilde{D}$.

(3) Use the two-site DMRG algorithm~\cite{AnnalPhysSchollwock} to optimize the MPS obtained in step (2) with respect to the target Hamiltonian [see Fig.~\ref{fig:intro}(c)], in which the bond dimension of the MPS is gradually increased from $\tilde{D}$ to $D$.

While the main steps are clear, a few subtleties turn out to be important for a successful implementation. Below we demonstrate these issues and benchmark the performance in the Kitaev honeycomb model.

\begin{figure}[tb]
\includegraphics[width=\linewidth]{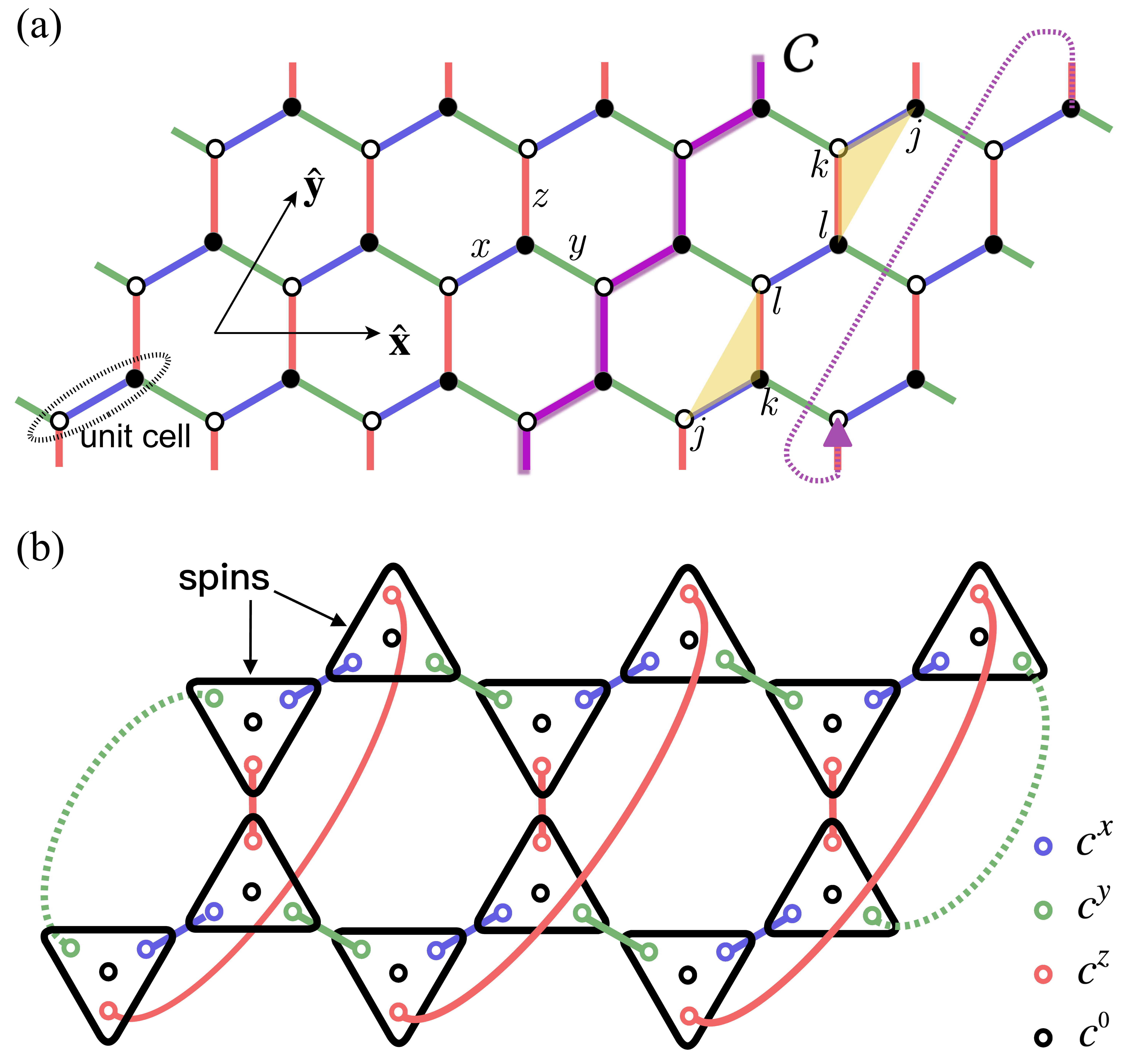}
\caption{(a) Kitaev honeycomb model on a cylinder geometry with two basis vectors $\hat{\mathbf{x}}$ and $\hat{\mathbf{y}}$, in which the $x$-boundary is open while the $y$-boundary is periodic. Black dots and white circles stand for A and B sublattice. $x$, $y$ and $z$ denote three types of bonds.
The three-spin interactions in Eq.~\eqref{eq:H-Kitaev} are defined on two types of triangles with vertexes $j$, $k$, and $l$. The purple zigzag line indicates a closed loop $\mathcal{C}$ along which the Wilson loop operator $W_y$ is defined, see Eq.~\eqref{eq:wilson-loop}. (b) Graphic representation of Kitaev's four-Majorana decomposition of spins. Solid bonds stand for the $\mathbb{Z}_2$ gauge field $u_{jk}$. Dash bonds emanating from the $x$-boundary indicate how to fix the unpaired boundary modes.}
\label{fig:KitaevHoneycomb}
\end{figure}

{\em Model.---} We first consider the Kitaev honeycomb model~\cite{Kitaev06} in the presence of three-spin interactions,
\begin{equation}
\mathcal{H}_3 = \sum_{a}\sum_{\langle jk\rangle\in a} J_{a} \sigma^a_j \sigma^a_k
+J_3 \sum_{\langle jkl \rangle\in\triangle} \sigma^x_{j}\sigma^y_{k}\sigma^{z}_{l},
\label{eq:H-Kitaev}
\end{equation}
where $\sigma_{j}^{a}$ $(a=x,y,z)$ are Pauli matrices, $\langle{}jk\rangle\in{}a$ denotes a nearest neighbor (NN) bond of type $a$ [see Fig.~\ref{fig:KitaevHoneycomb}(a)], and $\langle jkl \rangle\in\triangle$ refers to three sites around two types of triangles as indicated in Fig.~\ref{fig:KitaevHoneycomb}(a), as well as their translations to the whole lattice.

Following Kitaev's approach, we use the Majorana representation, $\sigma^{a}_j=ic^{a}_jc^{0}_j$, where $c^a$ ($c^0$) are so-called gauge (itinerant) Majorana fermions. This parton representation enlarges the Hilbert space and a local constraint $D_j\equiv c^{x}_jc^{y}_jc^{z}_jc^0_j=1$ has to be imposed to restore the physical Hilbert space of spin-1/2's. Under this representation, $\mathcal{H}_3$ becomes an effective Hamiltonian for Majorana partons,
\begin{equation}
H_{\mathrm{eff}} = -i \sum_a\sum_{\langle jk \rangle\in a}J_a{u}_{jk} c^0_j{}c^0_k -iJ_3\sum_{\langle jkl \rangle\in\triangle}u_{jk} {u}_{kl} c^0_j c^0_l,
\label{eq:Heff}
\end{equation}
where $u_{jk} \equiv ic_{j}^{a}c^{a}_{k}$ lives on an $a$-type bond.
Since $[H_{\mathrm{eff}},u_{jk}] = [u_{jk},u_{lm}] = 0$ for all different bonds, $u_{jk}$ are static $\mathbb{Z}_2$ gauge fields taking their eigenvalues $\pm 1$. When the gauge field configuration (denoted by $\{u\}$) is fixed, $H_{\mathrm{eff}}$ becomes a quadratic Hamiltonian of the itinerant Majorana fermion $c^{0}$, whose eigenstates can be written as $\vert\phi(\{u\})\rangle$. Together with the state of gauge Majorana fermions denoted by $\vert\{u\}\rangle$, the eigenstates of $H_{\mathrm{eff}}$ are given by
\begin{equation}
|\Psi_0\rangle = |\{u\}\rangle \otimes |\phi(\{u\})\rangle.
\label{eq:Psi0}
\end{equation}
These states are turned into (physical) eigenstates of the spin Hamiltonian $\mathcal{H}_3$ only after applying the Gutzwiller projection, i.e., $|\Psi_G\rangle=P_{G}|\Psi_0\rangle$ with $P_{G}\equiv\prod_{j}(1+D_j)/2$. Here the projection onto the singly occupied subspace can be revealed by combining Majoranas into complex fermions via $f_{j,\uparrow}=(c^x_j -ic^y_j)/2$ and $f_{j,\downarrow} = (c^z_j -ic^{0}_j)/2$, so that the local constraint becomes $\sum_{\sigma=\uparrow,\downarrow} f^\dag_{j,\sigma}f_{j,\sigma} = 1$. Accordingly, the ground state is achieved by determining the gauge field configuration $\{u\}$ in Eq.~(\ref{eq:Heff}) under which the resulting quadratic Hamiltonian of itinerant Majorana fermions has the lowest energy.

{\em MPO-MPS process.---} In correspondence with the common practice in DMRG, we adopt cylindrical boundary conditions, where the honeycomb lattice is embedded on a finite cylinder with $L_x$ ($L_y$) unit cells along the open (periodic) direction and a total number of $N=2L_xL_y$ sites. The Hamiltonian $\mathcal{H}_3$ now commutes with Wilson loop operators wrapping around the cylinder, e.g., $W_y = -\prod_{j \in \mathcal{C}} \sigma^y_j$ with $\mathcal{C}$ being a closed loop shown in Fig.~\ref{fig:KitaevHoneycomb}(a). The eigenvalue of $W_y$ is just the product of the static $\mathbb{Z}_2$ gauge fields along the loop,
\begin{equation}
W_y\vert \Psi_G\rangle = \Phi_y\vert\Psi_G\rangle,
\label{eq:wilson-loop}
\end{equation}
where $\Phi_y=\prod_{\langle jk \rangle \in \mathcal{C}} u_{jk} = \pm 1$.

The ground-state gauge configuration $\{u\}$ in $\Phi_y=1$ sector can be chosen as $u_{jk}=1$ for all bonds, while for $\Phi_y=-1$ sector it is achieved by setting $u_{jk}=-1$ for a row of $z$-bonds and $u_{jk}=1$ elsewhere~\cite{Kitaev06}. Here we have taken the convention that $j$ ($k$) belongs to A (B) sublattice [see Fig.~\ref{fig:KitaevHoneycomb}(a)].
However, it is worth emphasizing that, for each sector, there are still unpaired $c^y$ gauge Majorana fermions at the leftmost and rightmost boundaries [see Fig.~\ref{fig:KitaevHoneycomb}(a)], which do not enter into the Hamiltonian $H_{\mathrm{eff}}$ and thus lead to extra degeneracies. For the purpose of compressing the ground state into an MPS, we seek to minimize the entanglement, so we pair up these boundary gauge Majorana fermions [see Fig.~\ref{fig:KitaevHoneycomb}(b)] into complex fermions $f_{\langle\langle jl \rangle\rangle}\equiv(c^y_j - ic^y_l)/2$ and require that these boundary modes are unoccupied in the unprojected state $\vert\Psi_0\rangle$, i.e., $f_{\langle\langle jl \rangle\rangle}\vert\Psi_0\rangle = 0$ for all such boundary modes. Apparently, this manipulation is equivalent to adding suitable boundary terms in the spin Hamiltonian $\mathcal{H}_3$~\cite{appendix}, which is of great help in suppressing entanglement.

With these prescriptions, we are ready to convert the Gutzwiller projected state $\vert\Psi_G\rangle=P_G \vert\Psi_0\rangle$ into an MPS by noticing that $\vert\Psi_0\rangle = \prod_{m=1}^{2N}d^\dag_m\vert 0 \rangle$, where $\vert 0 \rangle$ is the vacuum of fermionic partons ($f_{j,\sigma}\vert 0\rangle = 0 \; \forall j,\sigma$) and $d^\dag_m$ are Bogoliubov-de Gennes (BdG) quasiparticle operators taking the form $d^\dag_m = \sum_{j=1}^{N}\sum_{\sigma=\uparrow,\downarrow} (U_{m,j\sigma} f^\dag_{j,\sigma} + V_{m,j\sigma} f_{j,\sigma})$ and satisfying $d^\dag_m \vert\Psi_0\rangle = 0$~\footnote{For gauge Majorana fermions, these BdG modes are trivially derived from the gauge choice $\{u\}$ and the fixing of the boundary modes. For itinerant Majorana fermions, the BdG modes are obtained by diagonalizing the quadratic Hamiltonian for $c^0$ under the fixed gauge choice, followed by Wannier localization~\cite{MPOMPS2}. More details can be found in the Supplemental Material.}. This form of $\vert\Psi_0\rangle$ is particularly suitable for utilizing the MPO-MPS method~\cite{MPOMPS2}, whose basic idea is summarized as follows [see Fig.~\ref{fig:intro}(b)]: (i) view each $d^\dag_m$ as an MPO and $\vert\Psi_0\rangle$ as a tensor network with $2N$ MPOs acting on a product state (parton vacuum); (ii) apply these MPOs successively (with a proper order) and compress the outcome in each intermediate step as an MPS with bond dimension up to $\tilde{D}$, which yields an MPS approximating $\vert \Psi_0\rangle$; (iii) apply the Gutzwiller projector $P_G$ to obtain $\vert \Psi_{\mathrm{MPS}}(\tilde{D})\rangle$, which is an MPS approximation of $\vert\Psi_G\rangle$. Further technical details are discussed in Ref.~\cite{appendix}.

At each intermediate step of the above MPO-MPS procedure, approximating the MPO-evolved MPS (with bond dimension $2\tilde{D}$) into an MPS (with bond dimension $\tilde{D}$) incurs a truncation error.
In order to estimate the accuracy of the final MPS, the accumulated truncation error is defined by
\begin{equation}
\epsilon_{\mathrm{trunc}}(\tilde{D})=1-\prod_{m=1}^{2N}F^{(m)}(\tilde{D}), \quad{}F^{(m)}(\tilde{D})=1-\sum^{2N}_{j=1}\epsilon^{(m)}_{j}(\tilde{D}),
\label{eq:trunc_error}
\end{equation}
where $\epsilon^{(m)}_{j}(\tilde{D})$ is the sum of the discarded squared singular values at the $j$-th bond of the $m$-th MPO-evolved MPS~\cite{AnnalPhysSchollwock}. Notice that $F^{(m)}(\tilde{D})$ is a rough estimate of the overlap between MPO-evolved MPS and truncated MPS in the $m$-th MPO-MPS step.

Since the Hamiltonian $\mathcal{H}_3$ in Eq.~\eqref{eq:H-Kitaev} is exactly solvable, we also quantify the errors, in both $\Phi_y = \pm 1$ sectors, by comparing the variational energy of the MPS $|\Psi_{\mathrm{MPS}}(\Phi_y,\tilde{D})\rangle$ with the exact ground-state energy $E_g(\Phi_y)$ via the relative energy deviation,
\begin{equation}
\delta E_g(\Phi_y,\tilde{D})=\frac{\langle\Psi_{\mathrm{MPS}}(\Phi_y,\tilde{D})\vert\mathcal{H}_{3}\vert \Psi_{\mathrm{MPS}}(\Phi_y,\tilde{D})\rangle - E_g(\Phi_y)}{|E_g(\Phi_y)|}.
\label{eq:dEg}
\end{equation}

\begin{table}[tb]
	\renewcommand\arraystretch{1.3}
	\setlength\tabcolsep{0.12cm}
	\begin{tabular}{c|c|c|c|c|c}
		\hline
		\hline
		& \multicolumn{4}{c|}{$J_x=1$} & $J_x=4$ \\
		\hline
		& $\tilde{D}$   &  $J_3=0$  & $J_3=0.1$ & $J_3=0.2$  & $J_3=0$\\
		\hline
		\multirow{6}{0.5cm}{$\epsilon_{\text{trunc}}$} &
		100	&
		$1.7\times10^{-1}$  &
		$9.2\times10^{-2}$ &
		$5.7\times10^{-2}$ &
		$1.1\times10^{-4}$  \\
		&	200 &
		$2.4\times10^{-2}$ &
		$1.0\times10^{-2}$ &
		$5.0\times10^{-3}$ &
		$1.0\times10^{-6}$ \\
		& {400}	&
		$2.5\times10^{-3}$ &
		$5.6\times10^{-4}$ &
		$2.4\times10^{-4}$ &
		$3.4\times10^{-7}$  \\
		& {600} &	
		$4.2\times10^{-4}$ &
		$8.0\times10^{-5}$ &
		$3.0\times10^{-5}$ &
		$3.4\times10^{-7}$  \\
		& {800} &
		$1.1\times10^{-4}$ &
		$1.9\times10^{-5}$ &
		$7.4\times10^{-6}$ &
		$3.4\times10^{-7}$  \\
		& {1000} &
		$3.4\times10^{-5}$ &
		$6.8\times10^{-6}$ &
		$2.9\times10^{-6}$ &
		$3.4\times10^{-7}$   \\		
		\hline
		\multirow{6}{0.5cm}{$\delta{}E_g$}  &
		100	&
		$1.3\times10^{-2}$  &
		$7.2\times10^{-3}$ &
		$3.9\times10^{-3}$ &
		$8.6\times10^{-5}$  \\
		& 200	&
		$1.1\times10^{-3}$  &
		$4.9\times10^{-4}$ &
		$1.8\times10^{-4}$ &
		$6.8\times10^{-8}$  \\
		& 400	&
		$8.8\times10^{-5}$  &
		$2.4\times10^{-5}$ &
		$9.2\times10^{-6}$ &
		$4.9\times10^{-8}$  \\		
		& 600	&
		$1.6\times10^{-5}$  &
		$4.0\times10^{-6}$ &
		$1.3\times10^{-6}$ &
		$4.9\times10^{-8}$  \\						
		& 800	&
		$4.4\times10^{-6}$  &
		$9.3\times10^{-7}$ &
		$3.3\times10^{-7}$ &
		$4.9\times10^{-8}$  \\		
		&1000&
		$1.6\times10^{-6}$  &
		$3.3\times10^{-7}$ &
		$1.3\times10^{-7}$ &
		$4.9\times10^{-8}$  \\		
		\hline
		\hline
	\end{tabular}
	\caption{The truncation error $\epsilon_{\text{trunc}}$ and the energy deviation $\delta E_g$ in the MPO-MPS process, which are defined in Eqs.~\eqref{eq:trunc_error} and \eqref{eq:dEg}, respectively. The MPO-MPS procedure is carried out for the Hamiltonian $\mathcal{H}_3$ with $J_y=J_z=1$, defined on a cylinder with $L_x \times L_y = 10 \times 4$ and in the $\Phi_y=-1$ sector.
}
\label{tab:epsilon}
\end{table}

To examine the accuracy of the MPO-MPS method, we compute the truncation error $\epsilon_{\text{trunc}}$ and the energy deviation $\delta{}E_g$ for the Hamiltonian $\mathcal{H}_{3}$ on a cylinder with $L_x\times L_y=10\times 4$ and in the sector $\Phi_y=-1$. We take $J_{y}=J_{z}=1$ and vary $J_{x}$ and $J_{3}$ to study both gapped and gapless { phases}. The results are summarized in Table~\ref{tab:epsilon}. For all these states, as increasing $\tilde{D}$, the truncation errors $\epsilon_{\text{trunc}}$ are significantly reduced. Nevertheless, the truncation error for the gapless case ($J_{x}=1$ and $J_{3}=0$) is clearly larger than {those in gapped phase}.
It is worth mentioning that, for the case with Abelian topological order ($J_{x}=4$ and $J_{3}=0$), the MPO-MPS procedure yields a highly accurate MPS approximation for the ground state. These results give a hint that good MPS approximations of Gutzwiller projected states could be obtained as long as the entanglement has been treated properly.

We are now in the position to perform DMRG optimization with initial MPSs prepared from Gutzwiller projected states. For this we consider the Hamiltonian $\mathcal{H}_3$ on a $L_x \times L_y = 6 \times 6$ cylinder and for the most challenging gapless case ($J_x=J_y=J_z=1$ and $J_3=0$)~\footnote{ In this case, the so-called ``loop-gas'' tensor network state in Ref.~\cite{lee2019} is an excellent trial wave function and would have been a nice initialization ansatz for DMRG. However, its MPS representation has a bond dimension $D=7^{L_y}$, which, for $L_y=6$, is beyond the computational capacity of DMRG.}. For this particular model, we obtain the MPS approximations of the ground states $|\Psi_{\mathrm{MPS}}(\Phi_y,D)\rangle$ in both $\Phi_{y}=\pm 1$ sectors. For comparison, we also randomly generate an MPS (with bond dimension $\tilde{D}$) and optimize it with the two-site DMRG until a converged MPS at bond dimension $D$ is obtained.

\begin{figure}[tb]
\includegraphics[width=\linewidth]{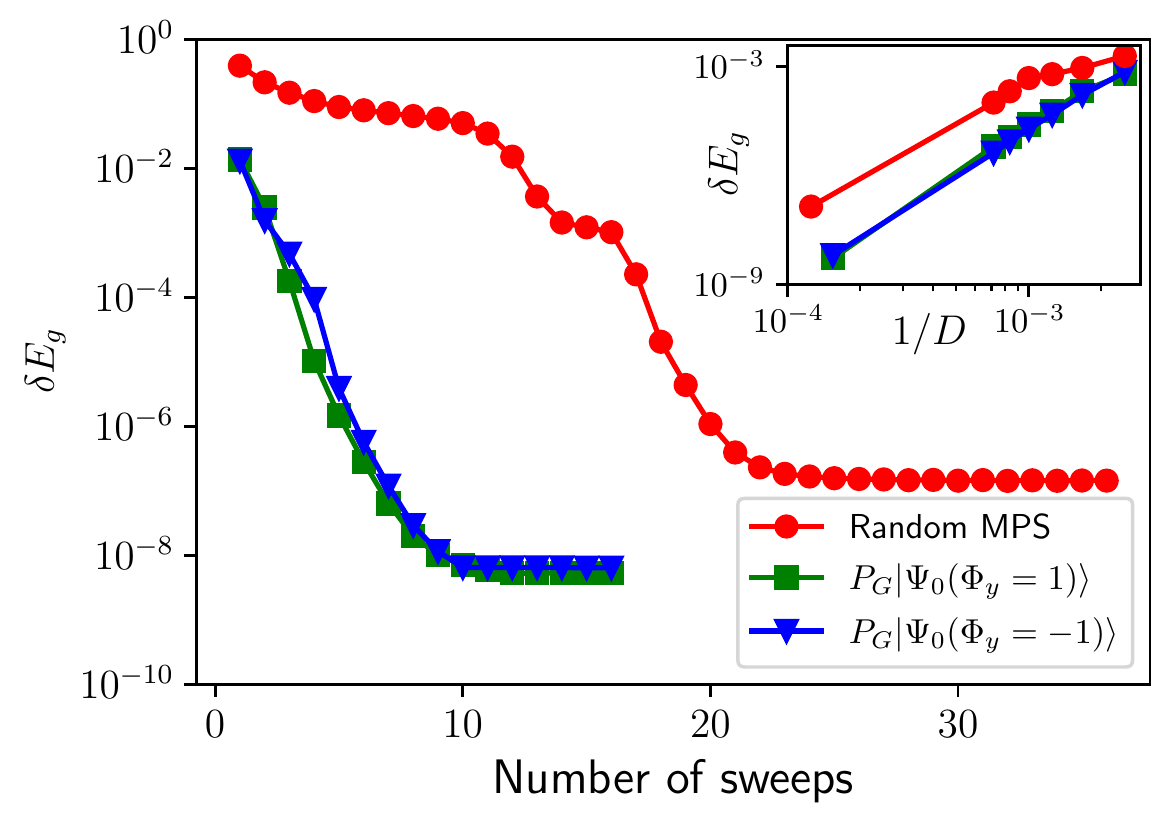}
%\caption{The relative energy deviations $\delta{}E_g$ [defined in Eq.~\eqref{eq:dEg}] versus number of sweeps in DMRG. The calculations are performed for the Hamiltonian $\mathcal{H}_3$ in Eq.~\eqref{eq:H-Kitaev} on an $L_x \times L_y =6 \times 6$ cylinder and with parameters $J_x=J_y=J_z=1$ and $J_3=0$. The bond dimension is chosen as $\tilde{D}=200$ for initial MPSs, and the first 5 DMRG sweeps are used to  gradually increase the bond dimension from $\tilde{D}$ to $D$. Red, green and blue lines stand for those with initial states of random MPS, $P_G|\Psi_0(\Phi_y=-1)\rangle$ and $P_G|\Psi_0(\Phi_y=1)\rangle$, respectively. Note that $\delta E_g$ initialized with a random MPS is measured from the ground-state energy in the $\Phi_y=-1$ sector. The final bond dimensions after DMRG sweeps are $D=600$, $1000$, $1400$ from top to bottom. Inset: $\delta E_g$ versus the inverse bond dimension $1/D$.}
\caption{{ The relative energy deviations $\delta{}E_g$ [defined in Eq.~\eqref{eq:dEg}] versus number of sweeps in DMRG. The calculations are performed for the Hamiltonian $\mathcal{H}_3$ in Eq.~\eqref{eq:H-Kitaev} on an $L_x \times L_y =6 \times 6$ cylinder and with parameters $J_x=J_y=J_z=1$ and $J_3=0$. The truncation errors are always kept to be smaller than $10^{-9}$ during DMRG optimization. Red, green and blue lines stand for those with initial states of random MPS, $P_G|\Psi_0(\Phi_y=-1)\rangle$ and $P_G|\Psi_0(\Phi_y=1)\rangle$, respectively. Note that $\delta E_g$ initialized with a random MPS is measured from the ground-state energy in the $\Phi_y=-1$ sector. The final bond dimensions after DMRG sweeps are $D=8000$ for random MPS and $D=6500$ for Gutzwiller ansatzes. Inset: $\delta E_g$ versus the inverse bond dimension $1/D$.}}
\label{fig:dEg_DMRG}
\end{figure}

The DMRG continues sweeping until $\delta{}E_{g}$ converges.
As illustrated in Fig.~\ref{fig:dEg_DMRG}, the relative energy deviation $\delta E_g$ is reduced by two orders of magnitude with Gutzwiller projected states $P_G|\Psi_0(\Phi_y=\pm{}1)\rangle$ being the initial ansatz.

In addition to the substantial improvement of the DMRG results, several remarks are in order:
(1) A relatively small bond dimension $\tilde{D}=200$ for the MPS prepared from $P_G|\Psi_0\rangle$ is sufficiently good to initialize the DMRG process, despite of a larger truncation error ($\epsilon_{\text{trunc}}\sim{}0.25$) in the MPO-MPS step. Meanwhile, the computational cost of preparing such MPS with $\tilde{D}=200$ is quite cheap.
(2) During the DMRG sweeps initialized with Gutzwiller projected states, the eigenvalue of the Wilson loop operator ($\Phi_y=\pm 1$) is preserved, i.e., the MPS stays in the respective sector. This is very useful for studying  topologically ordered states with topological degeneracy on the cylinder.
(3) { For the $6\times{}6$ cylinder, the DMRG initialized with a random MPS always converges to an MPS in $\Phi_y=-1$ sector.} However, exact results indicate that for a finite cylinder, the ground-state energy in $\Phi_y=-1$ sector is higher than that in the $\Phi_y=1$ sector. For instance, the energy difference on the $6 \times 6$ cylinder is given by $E_g(\Phi_y=-1)-E_g(\Phi_y=1)\approx 0.084$. This implies that the DMRG with a random initial ansatz gets stuck in a local minimum~\footnote{We have swept an unbiased set of random MPS up to 36 times, which is a sufficiently large number for DMRG optimization empirically. Starting from the 24th sweep, the (variational) ground-state energy does not decrease anymore and { becomes fluctuating, and the relative energy deviation $\delta{}E_g$ in the 36th sweep is almost identical to the one in the 24th sweep (the difference is about $3\times10^{-8}$)}, which clearly indicates that the randomly-initialized DMRG gets stuck in a local minimum.}.
(4) For the DMRG process initialized with a random MPS, $\delta E_g$ measured from $E_g(\Phi_y=-1)$ is still about two orders of magnitude larger than those initialized from the Gutzwiller projected state $P_G|\Psi_0(\Phi_y=-1)\rangle$. These clearly show that a properly chosen Gutzwiller projected state provides an ideal initialization ansatz for DMRG in two dimensions.

{\em Diagnosis of parton wave functions.---} The Gutzwiller-boosted DMRG is certainly applicable to generic models that do not have exact solutions. As a concrete example, we consider the Kitaev honeycomb model in an external magnetic field along the $[111]$ direction, defined by the Hamiltonian
\begin{equation}
\mathcal{H}= \sum_{\langle jk\rangle\in a} J_a \sigma^a_j \sigma^a_k
- h \sum_j (\sigma^x_j + \sigma^y_j + \sigma^z_j)
\label{eq:H1}
\end{equation}
with $J_x=J_y=J_z=1$. In this situation we shall focus on another function of our method, namely, diagnosing whether a Gutzwiller projected state captures the essential entanglement features of the actual ground state for a given Hamiltonian.

\begin{figure}[tb]
	\includegraphics[width=\linewidth]{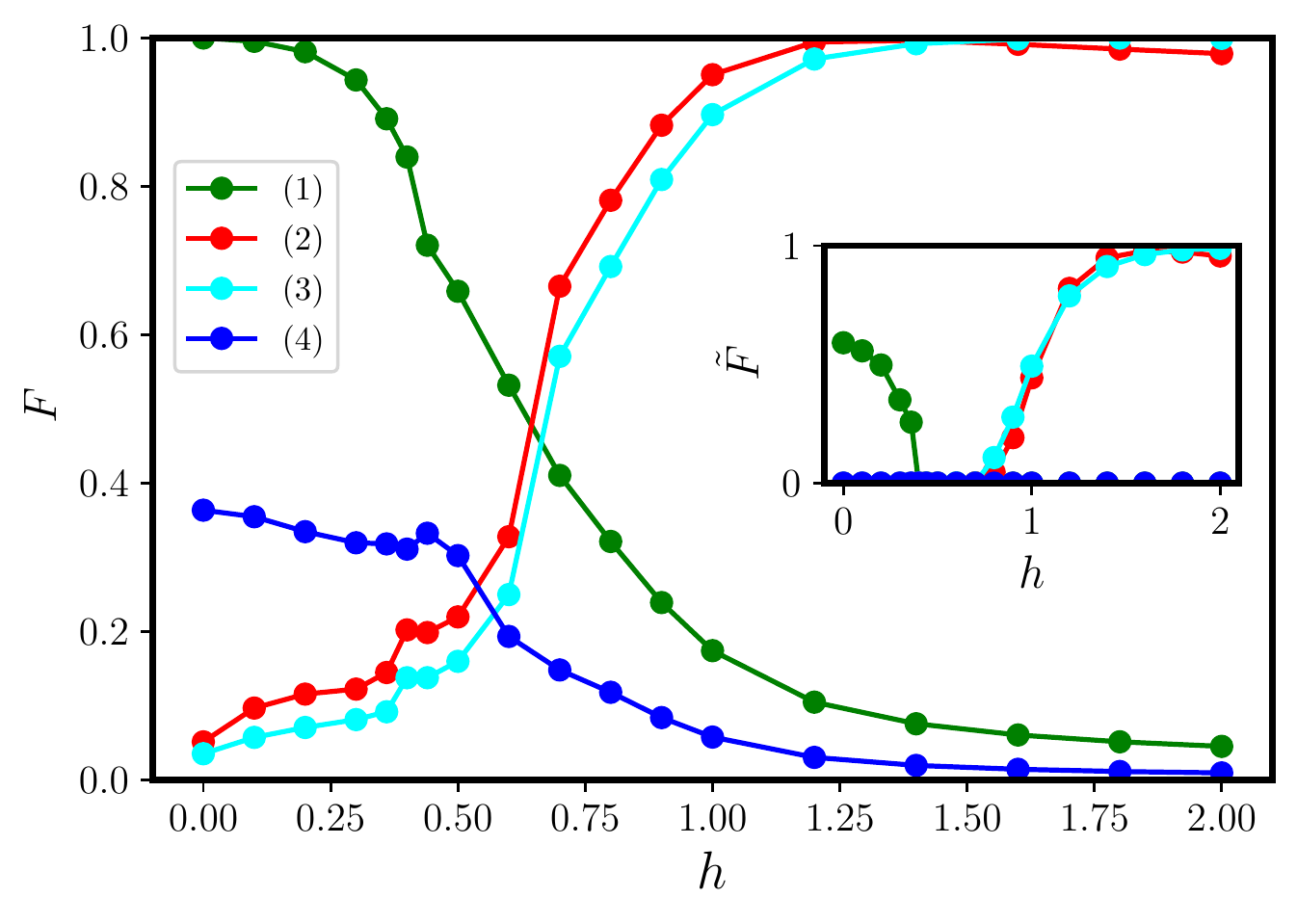}
	\caption{The fidelities $F$ as a function of the magnetic field $h$ for (1) the Kitaev's non-Abelian state (green), (2) fully polarized state (red), (3) partially polarized state (light blue), and (4) a $U(1)$ spin liquid state (dark blue). Further details on the parametrization of these states are given in Ref.~\cite{appendix}.
	The calculations are performed on an $L_x\times L_y = 10\times 4$ cylinder and the accumulated truncation errors in the MPO-MPS procedure are (1) $\epsilon_{\mathrm{trunc}}=0.005$, (2) $\epsilon_{\mathrm{trunc}}=0.12$,  (3) $\epsilon_{\mathrm{trunc}}=0.09$, and (4) $\epsilon_{\mathrm{trunc}}=0.01$ for these four states respectively.
	The DMRG calculations for the Hamiltonian in Eq.~\eqref{eq:H1} generate MPS with bond dimension $D=2400$ and truncation error $\epsilon_{\mathrm{DMRG}}\sim{}10^{-8}$ ($\epsilon_{\mathrm{DMRG}}\sim{}10^{-5}$ for $0.4\le{}h<0.7$).
	Inset: The wave-function fidelity $\tilde{F} = |\langle\Psi_\text{DMRG}|P_G|\Psi_\text{Parton}\rangle|$. }
	\label{fig:fidelity}
\end{figure}

In order to diagnose the quality of a Gutzwiller projected parton wave function $P_G|\Psi_\text{Parton}\rangle$, we utilize the fidelity defined by~\cite{McCulloch08}
\begin{equation}
F=\text{Tr}\left[\sqrt{\rho^{1/2}_{\text{D}}\rho^{}_{\text{G}}\rho^{1/2}_{\text{D}}}\right],
\label{eq:fidelity}
\end{equation}
where $\rho^{}_{\text{G}}$ and $\rho^{}_{\text{D}}$, being two reduced density matrices for a column of $2L_y$ sites in the middle of the cylinder, correspond to the Gutzwiller ansatz $P_G|\Psi_\text{Parton}\rangle$ and { the variational ground state $|\Psi_\text{DMRG}\rangle$ obtained by DMRG}, respectively.
This fidelity measures how close the bulk parts of two wave functions are, while the boundary effects due to the cylindrical geometry are precluded as much as possible. For comparison, we also evaluate the wave-function fidelity $\tilde{F} = |\langle\Psi_\text{DMRG}|P_G|\Psi_\text{Parton}\rangle|$.

While several parton constructions have been suggested for the Hamiltonian in Eq.~\eqref{eq:H1} (see, e.g., Refs.~\cite{Kitaev06,YMLu2018,QHWang2019}), we shall restrict ourselves to four classes of Gutzwiller ansatzes: (1) Kitaev's non-Abelian state with Chern number $C=1$; (2) fully polarized state with Chern number $C=0$; (3) partially polarized state with Chern number $C=1$; and (4) $U(1)$ spin liquid state with a spinon Fermi surface~\cite{YMLu2018}. Further details of these states can be found in the Supplemental Material~\cite{appendix}. The fidelities $F$ and $\tilde{F}$ between these Gutzwiller ansatzes and the ground state of the Hamiltonian in Eq.~\eqref{eq:H1} are shown in Fig.~\ref{fig:fidelity}, where the reference ground state is obtained by DMRG initialized with random MPSs.
It is seen that state (1) agrees well with the DMRG-obtained ground state at small $h$ ($0<h<0.35$), while both states (2) and (3) coincide with the ground state at large $h$ ($h>1.25$). For the whole region of $h$, the $U(1)$ spin liquid state (4) has negligible wave-function fidelity $\tilde{F}$, although the corresponding reduced-density-matrix fidelity $F$ is finite. It is worth noting that the two Gutzwiller projected states (2) and (3) have a large overlap with each other, although {their corresponding (unprojected) parton states carry different Chern numbers $C=0$ and $C=1$, respectively. For an intermediate magnetic field $h$ ($0.35<h<1$), we have observed that the DMRG \emph{cannot} be boosted by any of the four Gutzwiller ansatzes. This implies that none of these ansatzes describes actual ground states well.}
 	
{\em Summary.---} To summarize, we have devised a method to boost the performance of DMRG in two dimensions by using Gutzwiller projected states as the initialization ansatz. With the extensive benchmarks on the Kitaev honeycomb model, our method has shown clear advantages that { with suitably chosen Gutzwiller ansatz}, local minima are circumvented and much more accurate results are obtained with no extra computational costs. For topological states, the DMRG calculations initialized with Gutzwiller ansatz can preserve topological sectors, which is a very nice property for further characterizing the topological order~\cite{zhang2012,cincio2013,tu2013b,zaletel2013}. Our method also provides a diagnosis tool for analyzing the quality of Gutzwiller ansatz for a given Hamiltonian. Actually, a number of important strongly correlated systems have elusive ground states, albeit many parton wave function proposals are already available (e.g., spin-1/2 kagome Heisenberg antiferromagnet~\cite{yan2011,depenbrock2012,HJLiao2017,YCHe2017,YRan2007,Iqbal2013,TLi2018}). It would be interesting to revisit these problems armed with our new method.

{ Note added: After the submission of this work, we are aware of related works~\cite{petrica2020,aghaei2020} reporting results on converting Gutzwiller projected wave functions into MPSs and/or using them to initialize DMRG calculations.}

{\em Acknowledgement.---}
We thank Qiang-Hua Wang, Yang Qi, Hong Yao, Ying-Hai Wu, Urban Seifert, Yuan Wan and Zheng Zhu for helpful discussions. This work is supported in part by National Natural Science Foundation of China (No. 11774306), National Key Research and Development Program of China (No. 2016YFA0300202), the Strategic Priority Research Program of Chinese Academy of Sciences (No. XDB28000000) and the DFG through project A06 of SFB 1143 (project-id 247310070).

\bibliography{mpomps}

\begin{thebibliography}{50}%
\makeatletter
\providecommand \@ifxundefined [1]{%
 \@ifx{#1\undefined}
}%
\providecommand \@ifnum [1]{%
 \ifnum #1\expandafter \@firstoftwo
 \else \expandafter \@secondoftwo
 \fi
}%
\providecommand \@ifx [1]{%
 \ifx #1\expandafter \@firstoftwo
 \else \expandafter \@secondoftwo
 \fi
}%
\providecommand \natexlab [1]{#1}%
\providecommand \enquote  [1]{``#1''}%
\providecommand \bibnamefont  [1]{#1}%
\providecommand \bibfnamefont [1]{#1}%
\providecommand \citenamefont [1]{#1}%
\providecommand \href@noop [0]{\@secondoftwo}%
\providecommand \href [0]{\begingroup \@sanitize@url \@href}%
\providecommand \@href[1]{\@@startlink{#1}\@@href}%
\providecommand \@@href[1]{\endgroup#1\@@endlink}%
\providecommand \@sanitize@url [0]{\catcode `\\12\catcode `\$12\catcode
  `\&12\catcode `\#12\catcode `\^12\catcode `\_12\catcode `\%12\relax}%
\providecommand \@@startlink[1]{}%
\providecommand \@@endlink[0]{}%
\providecommand \url  [0]{\begingroup\@sanitize@url \@url }%
\providecommand \@url [1]{\endgroup\@href {#1}{\urlprefix }}%
\providecommand \urlprefix  [0]{URL }%
\providecommand \Eprint [0]{\href }%
\providecommand \doibase [0]{http://dx.doi.org/}%
\providecommand \selectlanguage [0]{\@gobble}%
\providecommand \bibinfo  [0]{\@secondoftwo}%
\providecommand \bibfield  [0]{\@secondoftwo}%
\providecommand \translation [1]{[#1]}%
\providecommand \BibitemOpen [0]{}%
\providecommand \bibitemStop [0]{}%
\providecommand \bibitemNoStop [0]{.\EOS\space}%
\providecommand \EOS [0]{\spacefactor3000\relax}%
\providecommand \BibitemShut  [1]{\csname bibitem#1\endcsname}%
\let\auto@bib@innerbib\@empty
%</preamble>
\bibitem [{\citenamefont {White}(1992)}]{white1992}%
  \BibitemOpen
  \bibfield  {author} {\bibinfo {author} {\bibfnamefont {S.~R.}\ \bibnamefont
  {White}},\ }\href {\doibase 10.1103/PhysRevLett.69.2863} {\bibfield
  {journal} {\bibinfo  {journal} {Phys. Rev. Lett.}\ }\textbf {\bibinfo
  {volume} {69}},\ \bibinfo {pages} {2863} (\bibinfo {year}
  {1992})}\BibitemShut {NoStop}%
\bibitem [{\citenamefont {White}(1993)}]{white1993}%
  \BibitemOpen
  \bibfield  {author} {\bibinfo {author} {\bibfnamefont {S.~R.}\ \bibnamefont
  {White}},\ }\href {\doibase 10.1103/PhysRevB.48.10345} {\bibfield  {journal}
  {\bibinfo  {journal} {Phys. Rev. B}\ }\textbf {\bibinfo {volume} {48}},\
  \bibinfo {pages} {10345} (\bibinfo {year} {1993})}\BibitemShut {NoStop}%
\bibitem [{\citenamefont {Schollw\"ock}(2005)}]{RMPSchollwock}%
  \BibitemOpen
  \bibfield  {author} {\bibinfo {author} {\bibfnamefont {U.}~\bibnamefont
  {Schollw\"ock}},\ }\href {\doibase 10.1103/RevModPhys.77.259} {\bibfield
  {journal} {\bibinfo  {journal} {Rev. Mod. Phys.}\ }\textbf {\bibinfo {volume}
  {77}},\ \bibinfo {pages} {259} (\bibinfo {year} {2005})}\BibitemShut
  {NoStop}%
\bibitem [{\citenamefont {Hallberg}(2006)}]{AdvPhyHallberg}%
  \BibitemOpen
  \bibfield  {author} {\bibinfo {author} {\bibfnamefont {K.~A.}\ \bibnamefont
  {Hallberg}},\ }\href {\doibase 10.1080/00018730600766432} {\bibfield
  {journal} {\bibinfo  {journal} {Adv. Phys.}\ }\textbf {\bibinfo {volume}
  {55}},\ \bibinfo {pages} {477} (\bibinfo {year} {2006})}\BibitemShut
  {NoStop}%
\bibitem [{\citenamefont {Schollw\"ock}(2011)}]{AnnalPhysSchollwock}%
  \BibitemOpen
  \bibfield  {author} {\bibinfo {author} {\bibfnamefont {U.}~\bibnamefont
  {Schollw\"ock}},\ }\href {\doibase https://doi.org/10.1016/j.aop.2010.09.012}
  {\bibfield  {journal} {\bibinfo  {journal} {Ann. Phys.}\ }\textbf {\bibinfo
  {volume} {326}},\ \bibinfo {pages} {96 } (\bibinfo {year}
  {2011})}\BibitemShut {NoStop}%
\bibitem [{\citenamefont {\"Ostlund}\ and\ \citenamefont
  {Rommer}(1995)}]{DMRG2MPS_1}%
  \BibitemOpen
  \bibfield  {author} {\bibinfo {author} {\bibfnamefont {S.}~\bibnamefont
  {\"Ostlund}}\ and\ \bibinfo {author} {\bibfnamefont {S.}~\bibnamefont
  {Rommer}},\ }\href {\doibase 10.1103/PhysRevLett.75.3537} {\bibfield
  {journal} {\bibinfo  {journal} {Phys. Rev. Lett.}\ }\textbf {\bibinfo
  {volume} {75}},\ \bibinfo {pages} {3537} (\bibinfo {year}
  {1995})}\BibitemShut {NoStop}%
\bibitem [{\citenamefont {Dukelsky}\ \emph {et~al.}(1998)\citenamefont
  {Dukelsky}, \citenamefont {Mart{\'{\i}}n-Delgado}, \citenamefont {Nishino},\
  and\ \citenamefont {Sierra}}]{DMRG2MPS_2}%
  \BibitemOpen
  \bibfield  {author} {\bibinfo {author} {\bibfnamefont {J.}~\bibnamefont
  {Dukelsky}}, \bibinfo {author} {\bibfnamefont {M.~A.}\ \bibnamefont
  {Mart{\'{\i}}n-Delgado}}, \bibinfo {author} {\bibfnamefont {T.}~\bibnamefont
  {Nishino}}, \ and\ \bibinfo {author} {\bibfnamefont {G.}~\bibnamefont
  {Sierra}},\ }\href {\doibase 10.1209/epl/i1998-00381-x} {\bibfield  {journal}
  {\bibinfo  {journal} {Europhys. Lett.}\ }\textbf {\bibinfo {volume} {43}},\
  \bibinfo {pages} {457} (\bibinfo {year} {1998})}\BibitemShut {NoStop}%
\bibitem [{\citenamefont {Verstraete}\ and\ \citenamefont
  {Cirac}(2006)}]{verstraete2006}%
  \BibitemOpen
  \bibfield  {author} {\bibinfo {author} {\bibfnamefont {F.}~\bibnamefont
  {Verstraete}}\ and\ \bibinfo {author} {\bibfnamefont {J.~I.}\ \bibnamefont
  {Cirac}},\ }\href {\doibase 10.1103/PhysRevB.73.094423} {\bibfield  {journal}
  {\bibinfo  {journal} {Phys. Rev. B}\ }\textbf {\bibinfo {volume} {73}},\
  \bibinfo {pages} {094423} (\bibinfo {year} {2006})}\BibitemShut {NoStop}%
\bibitem [{\citenamefont {Hastings}(2007)}]{hastings2007}%
  \BibitemOpen
  \bibfield  {author} {\bibinfo {author} {\bibfnamefont {M.~B.}\ \bibnamefont
  {Hastings}},\ }\href {https://doi.org/10.1088/1742-5468/2007/08/P08024}
  {\bibfield  {journal} {\bibinfo  {journal} {J. Stat. Mech.}\ }\textbf
  {\bibinfo {volume} {2007}},\ \bibinfo {pages} {P08024} (\bibinfo {year}
  {2007})}\BibitemShut {NoStop}%
\bibitem [{\citenamefont {Stoudenmire}\ and\ \citenamefont
  {White}(2012)}]{stoudenmire2012}%
  \BibitemOpen
  \bibfield  {author} {\bibinfo {author} {\bibfnamefont {E.}~\bibnamefont
  {Stoudenmire}}\ and\ \bibinfo {author} {\bibfnamefont {S.~R.}\ \bibnamefont
  {White}},\ }\href {\doibase 10.1146/annurev-conmatphys-020911-125018}
  {\bibfield  {journal} {\bibinfo  {journal} {Annu. Rev. Condens. Matter
  Phys.}\ }\textbf {\bibinfo {volume} {3}},\ \bibinfo {pages} {111} (\bibinfo
  {year} {2012})}\BibitemShut {NoStop}%
\bibitem [{\citenamefont {Vidal}\ \emph {et~al.}(2003)\citenamefont {Vidal},
  \citenamefont {Latorre}, \citenamefont {Rico},\ and\ \citenamefont
  {Kitaev}}]{entangle2D1}%
  \BibitemOpen
  \bibfield  {author} {\bibinfo {author} {\bibfnamefont {G.}~\bibnamefont
  {Vidal}}, \bibinfo {author} {\bibfnamefont {J.~I.}\ \bibnamefont {Latorre}},
  \bibinfo {author} {\bibfnamefont {E.}~\bibnamefont {Rico}}, \ and\ \bibinfo
  {author} {\bibfnamefont {A.}~\bibnamefont {Kitaev}},\ }\href {\doibase
  10.1103/PhysRevLett.90.227902} {\bibfield  {journal} {\bibinfo  {journal}
  {Phys. Rev. Lett.}\ }\textbf {\bibinfo {volume} {90}},\ \bibinfo {pages}
  {227902} (\bibinfo {year} {2003})}\BibitemShut {NoStop}%
\bibitem [{\citenamefont {Latorre}\ \emph {et~al.}(2004)\citenamefont
  {Latorre}, \citenamefont {Rico},\ and\ \citenamefont {Vidal}}]{entangle2D2}%
  \BibitemOpen
  \bibfield  {author} {\bibinfo {author} {\bibfnamefont {J.~I.}\ \bibnamefont
  {Latorre}}, \bibinfo {author} {\bibfnamefont {E.}~\bibnamefont {Rico}}, \
  and\ \bibinfo {author} {\bibfnamefont {G.}~\bibnamefont {Vidal}},\ }\href
  {\doibase https://doi.org/10.26421/QIC4.1} {\bibfield  {journal} {\bibinfo
  {journal} {Quant. Inf. Comput.}\ }\textbf {\bibinfo {volume} {4}},\ \bibinfo
  {pages} {48} (\bibinfo {year} {2004})}\BibitemShut {NoStop}%
\bibitem [{\citenamefont {Eisert}\ \emph {et~al.}(2010)\citenamefont {Eisert},
  \citenamefont {Cramer},\ and\ \citenamefont {Plenio}}]{RMPAreaLaws}%
  \BibitemOpen
  \bibfield  {author} {\bibinfo {author} {\bibfnamefont {J.}~\bibnamefont
  {Eisert}}, \bibinfo {author} {\bibfnamefont {M.}~\bibnamefont {Cramer}}, \
  and\ \bibinfo {author} {\bibfnamefont {M.~B.}\ \bibnamefont {Plenio}},\
  }\href {\doibase 10.1103/RevModPhys.82.277} {\bibfield  {journal} {\bibinfo
  {journal} {Rev. Mod. Phys.}\ }\textbf {\bibinfo {volume} {82}},\ \bibinfo
  {pages} {277} (\bibinfo {year} {2010})}\BibitemShut {NoStop}%
\bibitem [{\citenamefont {Gros}(1989)}]{Gros89}%
  \BibitemOpen
  \bibfield  {author} {\bibinfo {author} {\bibfnamefont {C.}~\bibnamefont
  {Gros}},\ }\href {\doibase http://dx.doi.org/10.1016/0003-4916(89)90077-8}
  {\bibfield  {journal} {\bibinfo  {journal} {Ann. Phys. (N.Y.)}\ }\textbf
  {\bibinfo {volume} {189}},\ \bibinfo {pages} {53 } (\bibinfo {year}
  {1989})}\BibitemShut {NoStop}%
\bibitem [{\citenamefont {Anderson}\ \emph {et~al.}(2004)\citenamefont
  {Anderson}, \citenamefont {Lee}, \citenamefont {Randeria}, \citenamefont
  {Rice}, \citenamefont {Trivedi},\ and\ \citenamefont {Zhang}}]{anderson04}%
  \BibitemOpen
  \bibfield  {author} {\bibinfo {author} {\bibfnamefont {P.~W.}\ \bibnamefont
  {Anderson}}, \bibinfo {author} {\bibfnamefont {P.~A.}\ \bibnamefont {Lee}},
  \bibinfo {author} {\bibfnamefont {M.}~\bibnamefont {Randeria}}, \bibinfo
  {author} {\bibfnamefont {T.~M.}\ \bibnamefont {Rice}}, \bibinfo {author}
  {\bibfnamefont {N.}~\bibnamefont {Trivedi}}, \ and\ \bibinfo {author}
  {\bibfnamefont {F.~C.}\ \bibnamefont {Zhang}},\ }\href {\doibase
  10.1088/0953-8984/16/24/r02} {\bibfield  {journal} {\bibinfo  {journal} {J.
  Phys.: Condens. Matter}\ }\textbf {\bibinfo {volume} {16}},\ \bibinfo {pages}
  {R755} (\bibinfo {year} {2004})}\BibitemShut {NoStop}%
\bibitem [{\citenamefont {Lee}\ \emph {et~al.}(2006)\citenamefont {Lee},
  \citenamefont {Nagaosa},\ and\ \citenamefont {Wen}}]{rmp06}%
  \BibitemOpen
  \bibfield  {author} {\bibinfo {author} {\bibfnamefont {P.~A.}\ \bibnamefont
  {Lee}}, \bibinfo {author} {\bibfnamefont {N.}~\bibnamefont {Nagaosa}}, \ and\
  \bibinfo {author} {\bibfnamefont {X.-G.}\ \bibnamefont {Wen}},\ }\href
  {\doibase 10.1103/RevModPhys.78.17} {\bibfield  {journal} {\bibinfo
  {journal} {Rev. Mod. Phys.}\ }\textbf {\bibinfo {volume} {78}},\ \bibinfo
  {pages} {17} (\bibinfo {year} {2006})}\BibitemShut {NoStop}%
\bibitem [{\citenamefont {Zhou}\ \emph {et~al.}(2017)\citenamefont {Zhou},
  \citenamefont {Kanoda},\ and\ \citenamefont {Ng}}]{QSLRMP}%
  \BibitemOpen
  \bibfield  {author} {\bibinfo {author} {\bibfnamefont {Y.}~\bibnamefont
  {Zhou}}, \bibinfo {author} {\bibfnamefont {K.}~\bibnamefont {Kanoda}}, \ and\
  \bibinfo {author} {\bibfnamefont {T.-K.}\ \bibnamefont {Ng}},\ }\href
  {\doibase 10.1103/RevModPhys.89.025003} {\bibfield  {journal} {\bibinfo
  {journal} {Rev. Mod. Phys.}\ }\textbf {\bibinfo {volume} {89}},\ \bibinfo
  {pages} {025003} (\bibinfo {year} {2017})}\BibitemShut {NoStop}%
\bibitem [{\citenamefont {Jin}\ \emph {et~al.}(2020)\citenamefont {Jin},
  \citenamefont {Tu},\ and\ \citenamefont {Zhou}}]{MPOMPS2}%
  \BibitemOpen
  \bibfield  {author} {\bibinfo {author} {\bibfnamefont {H.-K.}\ \bibnamefont
  {Jin}}, \bibinfo {author} {\bibfnamefont {H.-H.}\ \bibnamefont {Tu}}, \ and\
  \bibinfo {author} {\bibfnamefont {Y.}~\bibnamefont {Zhou}},\ }\href {\doibase
  10.1103/PhysRevB.101.165135} {\bibfield  {journal} {\bibinfo  {journal}
  {Phys. Rev. B}\ }\textbf {\bibinfo {volume} {101}},\ \bibinfo {pages}
  {165135} (\bibinfo {year} {2020})}\BibitemShut {NoStop}%
\bibitem [{\citenamefont {Wu}\ \emph {et~al.}(2020)\citenamefont {Wu},
  \citenamefont {Wang},\ and\ \citenamefont {Tu}}]{MPOMPS1}%
  \BibitemOpen
  \bibfield  {author} {\bibinfo {author} {\bibfnamefont {Y.-H.}\ \bibnamefont
  {Wu}}, \bibinfo {author} {\bibfnamefont {L.}~\bibnamefont {Wang}}, \ and\
  \bibinfo {author} {\bibfnamefont {H.-H.}\ \bibnamefont {Tu}},\ }\href
  {\doibase 10.1103/PhysRevLett.124.246401} {\bibfield  {journal} {\bibinfo
  {journal} {Phys. Rev. Lett.}\ }\textbf {\bibinfo {volume} {124}},\ \bibinfo
  {pages} {246401} (\bibinfo {year} {2020})}\BibitemShut {NoStop}%
\bibitem [{\citenamefont {Kitaev}(2006)}]{Kitaev06}%
  \BibitemOpen
  \bibfield  {author} {\bibinfo {author} {\bibfnamefont {A.}~\bibnamefont
  {Kitaev}},\ }\href {https://doi.org/10.1016/j.aop.2005.10.005} {\bibfield
  {journal} {\bibinfo  {journal} {Ann. Phys.}\ }\textbf {\bibinfo {volume}
  {321}},\ \bibinfo {pages} {2} (\bibinfo {year} {2006})}\BibitemShut {NoStop}%
\bibitem [{\citenamefont {Zhu}\ \emph {et~al.}(2018)\citenamefont {Zhu},
  \citenamefont {Kimchi}, \citenamefont {Sheng},\ and\ \citenamefont
  {Fu}}]{zhu2018}%
  \BibitemOpen
  \bibfield  {author} {\bibinfo {author} {\bibfnamefont {Z.}~\bibnamefont
  {Zhu}}, \bibinfo {author} {\bibfnamefont {I.}~\bibnamefont {Kimchi}},
  \bibinfo {author} {\bibfnamefont {D.~N.}\ \bibnamefont {Sheng}}, \ and\
  \bibinfo {author} {\bibfnamefont {L.}~\bibnamefont {Fu}},\ }\href {\doibase
  10.1103/PhysRevB.97.241110} {\bibfield  {journal} {\bibinfo  {journal} {Phys.
  Rev. B}\ }\textbf {\bibinfo {volume} {97}},\ \bibinfo {pages} {241110}
  (\bibinfo {year} {2018})}\BibitemShut {NoStop}%
\bibitem [{\citenamefont {Gohlke}\ \emph {et~al.}(2018)\citenamefont {Gohlke},
  \citenamefont {Moessner},\ and\ \citenamefont {Pollmann}}]{gohlke2018}%
  \BibitemOpen
  \bibfield  {author} {\bibinfo {author} {\bibfnamefont {M.}~\bibnamefont
  {Gohlke}}, \bibinfo {author} {\bibfnamefont {R.}~\bibnamefont {Moessner}}, \
  and\ \bibinfo {author} {\bibfnamefont {F.}~\bibnamefont {Pollmann}},\ }\href
  {\doibase 10.1103/PhysRevB.98.014418} {\bibfield  {journal} {\bibinfo
  {journal} {Phys. Rev. B}\ }\textbf {\bibinfo {volume} {98}},\ \bibinfo
  {pages} {014418} (\bibinfo {year} {2018})}\BibitemShut {NoStop}%
\bibitem [{\citenamefont {Hickey}\ and\ \citenamefont
  {Trebst}(2019)}]{hickey2019}%
  \BibitemOpen
  \bibfield  {author} {\bibinfo {author} {\bibfnamefont {C.}~\bibnamefont
  {Hickey}}\ and\ \bibinfo {author} {\bibfnamefont {S.}~\bibnamefont
  {Trebst}},\ }\href {https://doi.org/10.1038/s41467-019-08459-9} {\bibfield
  {journal} {\bibinfo  {journal} {Nat. Commun.}\ }\textbf {\bibinfo {volume}
  {10}},\ \bibinfo {pages} {1} (\bibinfo {year} {2019})}\BibitemShut {NoStop}%
\bibitem [{\citenamefont {Jiang}\ \emph {et~al.}(2018)\citenamefont {Jiang},
  \citenamefont {Wang}, \citenamefont {Huang},\ and\ \citenamefont
  {Lu}}]{YMLu2018}%
  \BibitemOpen
  \bibfield  {author} {\bibinfo {author} {\bibfnamefont {H.-C.}\ \bibnamefont
  {Jiang}}, \bibinfo {author} {\bibfnamefont {C.-Y.}\ \bibnamefont {Wang}},
  \bibinfo {author} {\bibfnamefont {B.}~\bibnamefont {Huang}}, \ and\ \bibinfo
  {author} {\bibfnamefont {Y.-M.}\ \bibnamefont {Lu}},\ }\href
  {https://arxiv.org/abs/1809.08247} {} (\bibinfo {year} {2018}),\ \Eprint
  {http://arxiv.org/abs/1809.08247} {arXiv:1809.08247} \BibitemShut {NoStop}%
\bibitem [{\citenamefont {Jiang}\ \emph {et~al.}(2020)\citenamefont {Jiang},
  \citenamefont {Liang}, \citenamefont {Chen}, \citenamefont {Qi},
  \citenamefont {Li},\ and\ \citenamefont {Wang}}]{QHWang2019}%
  \BibitemOpen
  \bibfield  {author} {\bibinfo {author} {\bibfnamefont {M.-H.}\ \bibnamefont
  {Jiang}}, \bibinfo {author} {\bibfnamefont {S.}~\bibnamefont {Liang}},
  \bibinfo {author} {\bibfnamefont {W.}~\bibnamefont {Chen}}, \bibinfo {author}
  {\bibfnamefont {Y.}~\bibnamefont {Qi}}, \bibinfo {author} {\bibfnamefont
  {J.-X.}\ \bibnamefont {Li}}, \ and\ \bibinfo {author} {\bibfnamefont {Q.-H.}\
  \bibnamefont {Wang}},\ }\href {\doibase 10.1103/PhysRevLett.125.177203}
  {\bibfield  {journal} {\bibinfo  {journal} {Phys. Rev. Lett.}\ }\textbf
  {\bibinfo {volume} {125}},\ \bibinfo {pages} {177203} (\bibinfo {year}
  {2020})}\BibitemShut {NoStop}%
\bibitem [{app()}]{appendix}%
  \BibitemOpen
  \href@noop {} {\bibinfo  {journal} {See the Supplemental Material for more
  details}\ }\BibitemShut {NoStop}%
\bibitem [{Note1()}]{Note1}%
  \BibitemOpen
\bibfield  {journal} {  }\bibinfo {note} {For gauge Majorana fermions, these
  BdG modes are trivially derived from the gauge choice $\protect \{u\protect
  \}$ and the fixing of the boundary modes. For itinerant Majorana fermions,
  the BdG modes are obtained by diagonalizing the quadratic Hamiltonian for
  $c^0$ under the fixed gauge choice, followed by Wannier localization~\cite
  {MPOMPS2}. More details can be found in the Supplemental
  Material.}\BibitemShut {Stop}%
\bibitem [{Note2()}]{Note2}%
  \BibitemOpen
  \bibinfo {note} {In this case, the so-called ``loop-gas'' tensor network
  state in Ref.~\cite {lee2019} is an excellent trial wave function and would
  have been a nice initialization ansatz for DMRG. However, its MPS
  representation has a bond dimension $D=7^{L_y}$, which, for $L_y=6$, is
  beyond the computational capacity of DMRG.}\BibitemShut {Stop}%
\bibitem [{Note3()}]{Note3}%
  \BibitemOpen
  \bibinfo {note} {We have swept an unbiased set of random MPS up to 36 times,
  which is a sufficiently large number for DMRG optimization empirically.
  Starting from the 24th sweep, the (variational) ground-state energy does not
  decrease anymore and { becomes fluctuating, and the relative energy deviation
  $\delta {}E_g$ in the 36th sweep is almost identical to the one in the 24th
  sweep (the difference is about $3\times 10^{-8}$)}, which clearly indicates
  that the randomly-initialized DMRG gets stuck in a local
  minimum.}\BibitemShut {Stop}%
\bibitem [{\citenamefont {McCulloch}(2008)}]{McCulloch08}%
  \BibitemOpen
  \bibfield  {author} {\bibinfo {author} {\bibfnamefont {I.}~\bibnamefont
  {McCulloch}},\ }\href {https://arxiv.org/abs/0804.2509} {} (\bibinfo {year}
  {2008}),\ \Eprint {http://arxiv.org/abs/0804.2509} {arXiv:0804.2509}
  \BibitemShut {NoStop}%
\bibitem [{\citenamefont {Zhang}\ \emph {et~al.}(2012)\citenamefont {Zhang},
  \citenamefont {Grover}, \citenamefont {Turner}, \citenamefont {Oshikawa},\
  and\ \citenamefont {Vishwanath}}]{zhang2012}%
  \BibitemOpen
  \bibfield  {author} {\bibinfo {author} {\bibfnamefont {Y.}~\bibnamefont
  {Zhang}}, \bibinfo {author} {\bibfnamefont {T.}~\bibnamefont {Grover}},
  \bibinfo {author} {\bibfnamefont {A.}~\bibnamefont {Turner}}, \bibinfo
  {author} {\bibfnamefont {M.}~\bibnamefont {Oshikawa}}, \ and\ \bibinfo
  {author} {\bibfnamefont {A.}~\bibnamefont {Vishwanath}},\ }\href {\doibase
  10.1103/PhysRevB.85.235151} {\bibfield  {journal} {\bibinfo  {journal} {Phys.
  Rev. B}\ }\textbf {\bibinfo {volume} {85}},\ \bibinfo {pages} {235151}
  (\bibinfo {year} {2012})}\BibitemShut {NoStop}%
\bibitem [{\citenamefont {Cincio}\ and\ \citenamefont
  {Vidal}(2013)}]{cincio2013}%
  \BibitemOpen
  \bibfield  {author} {\bibinfo {author} {\bibfnamefont {L.}~\bibnamefont
  {Cincio}}\ and\ \bibinfo {author} {\bibfnamefont {G.}~\bibnamefont {Vidal}},\
  }\href {\doibase 10.1103/PhysRevLett.110.067208} {\bibfield  {journal}
  {\bibinfo  {journal} {Phys. Rev. Lett.}\ }\textbf {\bibinfo {volume} {110}},\
  \bibinfo {pages} {067208} (\bibinfo {year} {2013})}\BibitemShut {NoStop}%
\bibitem [{\citenamefont {Tu}\ \emph {et~al.}(2013)\citenamefont {Tu},
  \citenamefont {Zhang},\ and\ \citenamefont {Qi}}]{tu2013b}%
  \BibitemOpen
  \bibfield  {author} {\bibinfo {author} {\bibfnamefont {H.-H.}\ \bibnamefont
  {Tu}}, \bibinfo {author} {\bibfnamefont {Y.}~\bibnamefont {Zhang}}, \ and\
  \bibinfo {author} {\bibfnamefont {X.-L.}\ \bibnamefont {Qi}},\ }\href
  {\doibase 10.1103/PhysRevB.88.195412} {\bibfield  {journal} {\bibinfo
  {journal} {Phys. Rev. B}\ }\textbf {\bibinfo {volume} {88}},\ \bibinfo
  {pages} {195412} (\bibinfo {year} {2013})}\BibitemShut {NoStop}%
\bibitem [{\citenamefont {Zaletel}\ \emph {et~al.}(2013)\citenamefont
  {Zaletel}, \citenamefont {Mong},\ and\ \citenamefont
  {Pollmann}}]{zaletel2013}%
  \BibitemOpen
  \bibfield  {author} {\bibinfo {author} {\bibfnamefont {M.~P.}\ \bibnamefont
  {Zaletel}}, \bibinfo {author} {\bibfnamefont {R.~S.~K.}\ \bibnamefont
  {Mong}}, \ and\ \bibinfo {author} {\bibfnamefont {F.}~\bibnamefont
  {Pollmann}},\ }\href {\doibase 10.1103/PhysRevLett.110.236801} {\bibfield
  {journal} {\bibinfo  {journal} {Phys. Rev. Lett.}\ }\textbf {\bibinfo
  {volume} {110}},\ \bibinfo {pages} {236801} (\bibinfo {year}
  {2013})}\BibitemShut {NoStop}%
\bibitem [{\citenamefont {Yan}\ \emph {et~al.}(2011)\citenamefont {Yan},
  \citenamefont {Huse},\ and\ \citenamefont {White}}]{yan2011}%
  \BibitemOpen
  \bibfield  {author} {\bibinfo {author} {\bibfnamefont {S.}~\bibnamefont
  {Yan}}, \bibinfo {author} {\bibfnamefont {D.~A.}\ \bibnamefont {Huse}}, \
  and\ \bibinfo {author} {\bibfnamefont {S.~R.}\ \bibnamefont {White}},\ }\href
  {https://science.sciencemag.org/content/332/6034/1173} {\bibfield  {journal}
  {\bibinfo  {journal} {Science}\ }\textbf {\bibinfo {volume} {332}},\ \bibinfo
  {pages} {1173} (\bibinfo {year} {2011})}\BibitemShut {NoStop}%
\bibitem [{\citenamefont {Depenbrock}\ \emph {et~al.}(2012)\citenamefont
  {Depenbrock}, \citenamefont {McCulloch},\ and\ \citenamefont
  {Schollw\"ock}}]{depenbrock2012}%
  \BibitemOpen
  \bibfield  {author} {\bibinfo {author} {\bibfnamefont {S.}~\bibnamefont
  {Depenbrock}}, \bibinfo {author} {\bibfnamefont {I.~P.}\ \bibnamefont
  {McCulloch}}, \ and\ \bibinfo {author} {\bibfnamefont {U.}~\bibnamefont
  {Schollw\"ock}},\ }\href {\doibase 10.1103/PhysRevLett.109.067201} {\bibfield
   {journal} {\bibinfo  {journal} {Phys. Rev. Lett.}\ }\textbf {\bibinfo
  {volume} {109}},\ \bibinfo {pages} {067201} (\bibinfo {year}
  {2012})}\BibitemShut {NoStop}%
\bibitem [{\citenamefont {Liao}\ \emph {et~al.}(2017)\citenamefont {Liao},
  \citenamefont {Xie}, \citenamefont {Chen}, \citenamefont {Liu}, \citenamefont
  {Xie}, \citenamefont {Huang}, \citenamefont {Normand},\ and\ \citenamefont
  {Xiang}}]{HJLiao2017}%
  \BibitemOpen
  \bibfield  {author} {\bibinfo {author} {\bibfnamefont {H.~J.}\ \bibnamefont
  {Liao}}, \bibinfo {author} {\bibfnamefont {Z.~Y.}\ \bibnamefont {Xie}},
  \bibinfo {author} {\bibfnamefont {J.}~\bibnamefont {Chen}}, \bibinfo {author}
  {\bibfnamefont {Z.~Y.}\ \bibnamefont {Liu}}, \bibinfo {author} {\bibfnamefont
  {H.~D.}\ \bibnamefont {Xie}}, \bibinfo {author} {\bibfnamefont {R.~Z.}\
  \bibnamefont {Huang}}, \bibinfo {author} {\bibfnamefont {B.}~\bibnamefont
  {Normand}}, \ and\ \bibinfo {author} {\bibfnamefont {T.}~\bibnamefont
  {Xiang}},\ }\href {\doibase 10.1103/PhysRevLett.118.137202} {\bibfield
  {journal} {\bibinfo  {journal} {Phys. Rev. Lett.}\ }\textbf {\bibinfo
  {volume} {118}},\ \bibinfo {pages} {137202} (\bibinfo {year}
  {2017})}\BibitemShut {NoStop}%
\bibitem [{\citenamefont {He}\ \emph {et~al.}(2017)\citenamefont {He},
  \citenamefont {Zaletel}, \citenamefont {Oshikawa},\ and\ \citenamefont
  {Pollmann}}]{YCHe2017}%
  \BibitemOpen
  \bibfield  {author} {\bibinfo {author} {\bibfnamefont {Y.-C.}\ \bibnamefont
  {He}}, \bibinfo {author} {\bibfnamefont {M.~P.}\ \bibnamefont {Zaletel}},
  \bibinfo {author} {\bibfnamefont {M.}~\bibnamefont {Oshikawa}}, \ and\
  \bibinfo {author} {\bibfnamefont {F.}~\bibnamefont {Pollmann}},\ }\href
  {\doibase 10.1103/PhysRevX.7.031020} {\bibfield  {journal} {\bibinfo
  {journal} {Phys. Rev. X}\ }\textbf {\bibinfo {volume} {7}},\ \bibinfo {pages}
  {031020} (\bibinfo {year} {2017})}\BibitemShut {NoStop}%
\bibitem [{\citenamefont {Ran}\ \emph {et~al.}(2007)\citenamefont {Ran},
  \citenamefont {Hermele}, \citenamefont {Lee},\ and\ \citenamefont
  {Wen}}]{YRan2007}%
  \BibitemOpen
  \bibfield  {author} {\bibinfo {author} {\bibfnamefont {Y.}~\bibnamefont
  {Ran}}, \bibinfo {author} {\bibfnamefont {M.}~\bibnamefont {Hermele}},
  \bibinfo {author} {\bibfnamefont {P.~A.}\ \bibnamefont {Lee}}, \ and\
  \bibinfo {author} {\bibfnamefont {X.-G.}\ \bibnamefont {Wen}},\ }\href
  {\doibase 10.1103/PhysRevLett.98.117205} {\bibfield  {journal} {\bibinfo
  {journal} {Phys. Rev. Lett.}\ }\textbf {\bibinfo {volume} {98}},\ \bibinfo
  {pages} {117205} (\bibinfo {year} {2007})}\BibitemShut {NoStop}%
\bibitem [{\citenamefont {Iqbal}\ \emph {et~al.}(2013)\citenamefont {Iqbal},
  \citenamefont {Becca}, \citenamefont {Sorella},\ and\ \citenamefont
  {Poilblanc}}]{Iqbal2013}%
  \BibitemOpen
  \bibfield  {author} {\bibinfo {author} {\bibfnamefont {Y.}~\bibnamefont
  {Iqbal}}, \bibinfo {author} {\bibfnamefont {F.}~\bibnamefont {Becca}},
  \bibinfo {author} {\bibfnamefont {S.}~\bibnamefont {Sorella}}, \ and\
  \bibinfo {author} {\bibfnamefont {D.}~\bibnamefont {Poilblanc}},\ }\href
  {\doibase 10.1103/PhysRevB.87.060405} {\bibfield  {journal} {\bibinfo
  {journal} {Phys. Rev. B}\ }\textbf {\bibinfo {volume} {87}},\ \bibinfo
  {pages} {060405} (\bibinfo {year} {2013})}\BibitemShut {NoStop}%
\bibitem [{\citenamefont {Li}(2018)}]{TLi2018}%
  \BibitemOpen
  \bibfield  {author} {\bibinfo {author} {\bibfnamefont {T.}~\bibnamefont
  {Li}},\ }\href {https://arxiv.org/abs/1807.09463} {} (\bibinfo {year}
  {2018}),\ \Eprint {http://arxiv.org/abs/1807.09463} {arXiv:1807.09463}
  \BibitemShut {NoStop}%
\bibitem [{\citenamefont {Petrica}\ \emph {et~al.}(2021)\citenamefont
  {Petrica}, \citenamefont {Zheng}, \citenamefont {Chan},\ and\ \citenamefont
  {Clark}}]{petrica2020}%
  \BibitemOpen
  \bibfield  {author} {\bibinfo {author} {\bibfnamefont {G.}~\bibnamefont
  {Petrica}}, \bibinfo {author} {\bibfnamefont {B.-X.}\ \bibnamefont {Zheng}},
  \bibinfo {author} {\bibfnamefont {G.~K.-L.}\ \bibnamefont {Chan}}, \ and\
  \bibinfo {author} {\bibfnamefont {B.~K.}\ \bibnamefont {Clark}},\ }\href
  {\doibase 10.1103/PhysRevB.103.125161} {\bibfield  {journal} {\bibinfo
  {journal} {Phys. Rev. B}\ }\textbf {\bibinfo {volume} {103}},\ \bibinfo
  {pages} {125161} (\bibinfo {year} {2021})}\BibitemShut {NoStop}%
\bibitem [{\citenamefont {Aghaei}\ \emph {et~al.}(2020)\citenamefont {Aghaei},
  \citenamefont {Bauer}, \citenamefont {Shtengel},\ and\ \citenamefont
  {Mishmash}}]{aghaei2020}%
  \BibitemOpen
  \bibfield  {author} {\bibinfo {author} {\bibfnamefont {A.~M.}\ \bibnamefont
  {Aghaei}}, \bibinfo {author} {\bibfnamefont {B.}~\bibnamefont {Bauer}},
  \bibinfo {author} {\bibfnamefont {K.}~\bibnamefont {Shtengel}}, \ and\
  \bibinfo {author} {\bibfnamefont {R.~V.}\ \bibnamefont {Mishmash}},\ }\href
  {https://arxiv.org/abs/2009.12435} {} (\bibinfo {year} {2020}),\ \Eprint
  {http://arxiv.org/abs/2009.12435} {arXiv:2009.12435} \BibitemShut {NoStop}%
\bibitem [{\citenamefont {Abrikosov}(1965)}]{Abrikosv}%
  \BibitemOpen
  \bibfield  {author} {\bibinfo {author} {\bibfnamefont {A.~A.}\ \bibnamefont
  {Abrikosov}},\ }\href {\doibase 10.1103/PhysicsPhysiqueFizika.2.61}
  {\bibfield  {journal} {\bibinfo  {journal} {Physics Physique Fizika}\
  }\textbf {\bibinfo {volume} {2}},\ \bibinfo {pages} {61} (\bibinfo {year}
  {1965})}\BibitemShut {NoStop}%
\bibitem [{\citenamefont {Burnell}\ and\ \citenamefont
  {Nayak}(2011)}]{Burnell_Nayak_11}%
  \BibitemOpen
  \bibfield  {author} {\bibinfo {author} {\bibfnamefont {F.~J.}\ \bibnamefont
  {Burnell}}\ and\ \bibinfo {author} {\bibfnamefont {C.}~\bibnamefont
  {Nayak}},\ }\href {\doibase 10.1103/PhysRevB.84.125125} {\bibfield  {journal}
  {\bibinfo  {journal} {Phys. Rev. B}\ }\textbf {\bibinfo {volume} {84}},\
  \bibinfo {pages} {125125} (\bibinfo {year} {2011})}\BibitemShut {NoStop}%
\bibitem [{\citenamefont {You}\ \emph {et~al.}(2012)\citenamefont {You},
  \citenamefont {Kimchi},\ and\ \citenamefont {Vishwanath}}]{you12}%
  \BibitemOpen
  \bibfield  {author} {\bibinfo {author} {\bibfnamefont {Y.-Z.}\ \bibnamefont
  {You}}, \bibinfo {author} {\bibfnamefont {I.}~\bibnamefont {Kimchi}}, \ and\
  \bibinfo {author} {\bibfnamefont {A.}~\bibnamefont {Vishwanath}},\ }\href
  {\doibase 10.1103/PhysRevB.86.085145} {\bibfield  {journal} {\bibinfo
  {journal} {Phys. Rev. B}\ }\textbf {\bibinfo {volume} {86}},\ \bibinfo
  {pages} {085145} (\bibinfo {year} {2012})}\BibitemShut {NoStop}%
\bibitem [{\citenamefont {Wen}(2002{\natexlab{a}})}]{PSG1}%
  \BibitemOpen
  \bibfield  {author} {\bibinfo {author} {\bibfnamefont {X.-G.}\ \bibnamefont
  {Wen}},\ }\href {\doibase https://doi.org/10.1016/S0375-9601(02)00808-3}
  {\bibfield  {journal} {\bibinfo  {journal} {Phys. Lett. A}\ }\textbf
  {\bibinfo {volume} {300}},\ \bibinfo {pages} {175 } (\bibinfo {year}
  {2002}{\natexlab{a}})}\BibitemShut {NoStop}%
\bibitem [{\citenamefont {Wen}(2002{\natexlab{b}})}]{PSG2}%
  \BibitemOpen
  \bibfield  {author} {\bibinfo {author} {\bibfnamefont {X.-G.}\ \bibnamefont
  {Wen}},\ }\href {\doibase 10.1103/PhysRevB.65.165113} {\bibfield  {journal}
  {\bibinfo  {journal} {Phys. Rev. B}\ }\textbf {\bibinfo {volume} {65}},\
  \bibinfo {pages} {165113} (\bibinfo {year} {2002}{\natexlab{b}})}\BibitemShut
  {NoStop}%
\bibitem [{\citenamefont {Zhou}\ and\ \citenamefont {Wen}(2002)}]{PSG3}%
  \BibitemOpen
  \bibfield  {author} {\bibinfo {author} {\bibfnamefont {Y.}~\bibnamefont
  {Zhou}}\ and\ \bibinfo {author} {\bibfnamefont {X.-G.}\ \bibnamefont {Wen}},\
  }\href {https://arxiv.org/abs/cond-mat/0210662} {\  (\bibinfo {year}
  {2002})},\ \Eprint {http://arxiv.org/abs/cond-mat/0210662}
  {arXiv:cond-mat/0210662} \BibitemShut {NoStop}%
\bibitem [{\citenamefont {Lee}\ \emph {et~al.}(2019)\citenamefont {Lee},
  \citenamefont {Kaneko}, \citenamefont {Okubo},\ and\ \citenamefont
  {Kawashima}}]{lee2019}%
  \BibitemOpen
  \bibfield  {author} {\bibinfo {author} {\bibfnamefont {H.-Y.}\ \bibnamefont
  {Lee}}, \bibinfo {author} {\bibfnamefont {R.}~\bibnamefont {Kaneko}},
  \bibinfo {author} {\bibfnamefont {T.}~\bibnamefont {Okubo}}, \ and\ \bibinfo
  {author} {\bibfnamefont {N.}~\bibnamefont {Kawashima}},\ }\href {\doibase
  10.1103/PhysRevLett.123.087203} {\bibfield  {journal} {\bibinfo  {journal}
  {Phys. Rev. Lett.}\ }\textbf {\bibinfo {volume} {123}},\ \bibinfo {pages}
  {087203} (\bibinfo {year} {2019})}\BibitemShut {NoStop}%
\end{thebibliography}%

\begin{widetext}
	
	\begin{center}
		{\bf Supplemental material for ``Density matrix renormalization group boosted by Gutzwiller projected wave functions"}
	\end{center}
	
	\setcounter{equation}{0}
	\setcounter{figure}{0}
	\setcounter{table}{0}
	
	\renewcommand{\theequation}{S\arabic{equation}}
	\renewcommand{\thefigure}{S\arabic{figure}}
	\renewcommand{\thetable}{S\Roman{table}}

	This Supplemental Material provides technique details: (1) the implementation of the MPO-MPS method, (2) numerical data for Fig.~3 in the main text,  (3) the four classes of Gutzwiller projected wave functions used in the main text,  {(4) discussion about $\{W_y,W_p\}$-fixed ansatz, and (5) entanglement spectra for Kitaev's non-Abelian states}.
	
	\section{The implementation of the MPO-MPS method}
	\label{sec:MPOMPS}
	
	In this section, we provide details on how to convert a Gutzwiller projected state $|\Psi_{G}\rangle=P_{G}|\Psi_{0}\rangle$ to an MPS by using the MPO-MPS method, where $|\Psi_0\rangle=|\{u\}\rangle\otimes{}|\phi(\{u\})\rangle$ is the unprojected ground state of the Kitaev honeycomb model that is defined in Eq.~(3) in the main text.
	
	\subsection{One-dimensional path}
	%{\bf{}One-dimensional path.---}
	In order to carry out the MPO-MPS procedure and perform the DMRG optimization, one should first define the ordering of lattice sites. This can be done by assigning an integer  $\tilde{j}=1,\cdots,N$ to each lattice site. There are two frequently used site-labeling schemes for an $L_x\times{}L_y$ honeycomb lattice on a cylinder as illustrated in Fig.~\ref{fig:quasi1d}. More explicitly, the lattice site belonging to unit cell $\mathbf{r}=\mathrm{r}_x{}\hat{\mathbf{x}}+\mathrm{r}_y\hat{\mathbf{y}}$ ($\mathrm{r}_x=1,\cdots,L_x$ and $\mathrm{r}_y=1,\cdots,L_y$) and sublattice A or B can be labeled by two different schemes as follows:\\
	(1) [see Fig.~\ref{fig:quasi1d}(a)]
	\begin{subequations}\label{eq:label12}
		\begin{equation}\label{eq:label1}
			\begin{split}
				&\tilde{j} = 2[(\mathrm{r}_x-1)L_y+\mathrm{r}_y-1]+\mathrm{C_{AB}},
			\end{split}
		\end{equation}
		or\\
		(2) [see Fig.~\ref{fig:quasi1d}(b)]
		\begin{equation}\label{eq:label2}
			\begin{split}
				&\begin{array}{ll}
					\tilde{j} = 2[(\mathrm{r}_x-1)L_y+\mathrm{r}_y-1]+\mathrm{C_{AB}},\quad &\text{for odd } r_x,\\
					\tilde{j} = 2(\mathrm{r}_xL_y-\mathrm{r}_y )+\mathrm{C^{\prime}_{AB}},\quad &\text{for even } r_x ,
				\end{array}
			\end{split}
		\end{equation}
		where $\mathrm{C_{AB}}=1\ (2)$ and $\mathrm{C^{\prime}_{ab}}=2\ (1)$ for A (B) sublattice.
	\end{subequations}
	
	\begin{figure}[tbhp]
		\includegraphics[width=17cm]{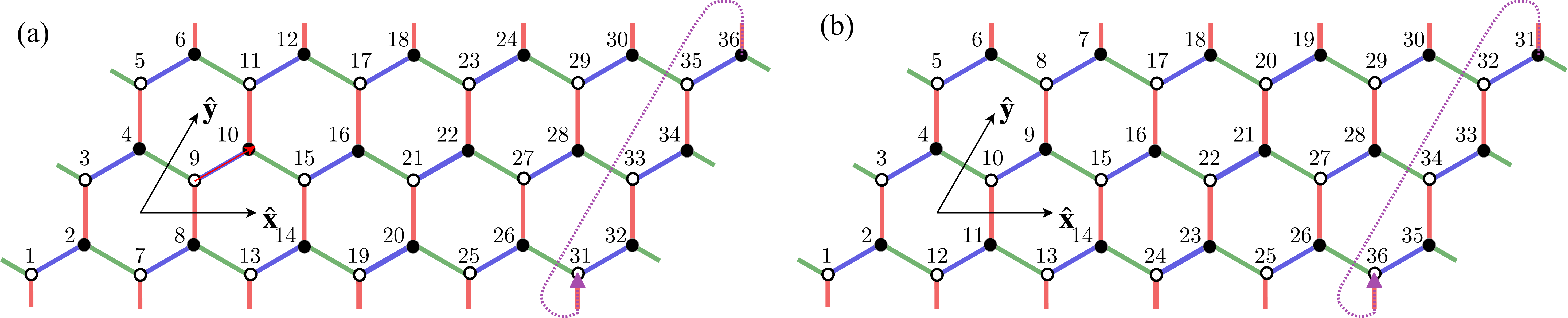}
		\caption{Schematics of two labeling schemes for a honeycomb lattice on an $L_x\times{}L_y=6\times{}3$ cylinder with basis vectors $\hat{\mathbf{x}}$ and $\hat{\mathbf{y}}$. (a) and (b) correspond to the labeling scheme given in Eq.~\eqref{eq:label1} and Eq.~\eqref{eq:label2}, respectively. With these two labeling schemes, one is able to carry out the MPO-MPS procedure and perform the DMRG calculation.}
		\label{fig:quasi1d}
	\end{figure}
	
	In the MPO-MPS procedure, we found that scheme (1) [defined in Eq.~\eqref{eq:label1} and shown in Fig.~\ref{fig:quasi1d} (a)] always gives rise to a smaller accumulated truncation error $\epsilon_{\text{trunc}}$ [see Eq.~(5) in the main text for its definition] than that obtained by scheme (2) [defined in  Eq.~\eqref{eq:label2} and shown in Fig.~\ref{fig:quasi1d} (b)].
	For instance, consider the Hamiltonian $\mathcal{H}_3$ with $J_x=J_y=J_z=1$ on an $L_x\times{}L_y=10\times{}4$ cylinder and in the $\Phi_y=-1$ sector, the truncation errors are given by $\epsilon_{\text{trunc}}(\tilde{D}=100)\approx{}0.09$ for scheme (1) and $\epsilon_{\text{trunc}}(\tilde{D}=100)\approx{}0.24$ for scheme (2), respectively. Thereby, throughout this work, we utilize scheme (1) to define the ordering of lattice sites and thus convert a cylinder into a 1D chain to apply the MPO-MPS method and perform DMRG calculations.
	
	\subsection{Complex fermion representation}
	%{\bf{}Complex fermion representation.---}
	To apply the MPO-MPS method in Refs.~[\onlinecite{MPOMPS1,MPOMPS2}], it is convenient to rewrite the Majorana ground state $|\Psi_0\rangle$ in terms of complex fermions~\cite{Abrikosv}. In the standard gauge theory approach to a quantum spin $S=1/2$ system~\cite{QSLRMP}, a complex Abrikosov fermion doublet $(\vec{f}_j)^\dagger\equiv{}(f^\dagger_{j,\uparrow}~f^\dagger_{j,\downarrow})$ is introduced at each site $j$ to represent spin operators,
	\begin{equation*}
		S^a_j = \frac{1}{2}(\vec{f}_j)^\dagger\sigma^a{}\vec{f}_j, \quad a=x,y,z.
	\end{equation*}
	With a certain $SU(2)$ gauge choice\cite{Burnell_Nayak_11,you12,QSLRMP}, complex Abrikosov fermions $f_{j,\uparrow(\downarrow)}$ are related to the four Majorana fermions as follows:
	\begin{equation}
		\begin{split}
			f_{j,\uparrow}=\frac{1}{2}\left(c^x_j-i c^y_j\right),\qquad
			f_{j,\downarrow}=\frac{1}{2}\left(c^z_j-ic^0_j\right), \label{eq:c2a}
		\end{split}
	\end{equation}
	Here it is easy to see that the local constraint $D_j=c_j^{x}c_j^{y}c_j^{z}c_j^{0}=1$ for Majorana fermions becomes the single-occupancy condition $f^\dagger_{j,\uparrow}f_{j,\uparrow}+f^\dagger_{j,\downarrow}f_{j,\downarrow}=1$ for complex fermions.

	\subsection{Bogoliubov-de-Gennes (BdG) formulation}
	%{\bf{}Bogoliubov-de-Gennes (BdG) formulation.---}
	Now we shall explain how to implement the MPO-MPS method with the help of a BdG Hamiltonian, as developed in Ref.~[\onlinecite{MPOMPS2}]. For simplicity, we shall focus on the $\Phi_y=-1$ sector. In accordance with Eq.~(3) in the main text, the unprojected ground state $|\Psi_0(\Phi_y=-1)\rangle$ is written as
	\begin{equation*}
		|\Psi_0(\Phi_y=-1)\rangle =|\{u\}(\Phi_y=-1)\rangle\otimes{}|\phi(\Phi_y=-1)\rangle,
	\end{equation*}
	where $|\{u\}(\Phi_y=-1)\rangle$ denotes the ground state of gauge Majorana fermions with  $\Phi_y=-1$ and $|\phi(\Phi_y=-1)\rangle$ the corresponding itinerant Majorana ground state. As mentioned in the main text, such a $\{u\}(\Phi_y=-1)$ configuration can be achieved by setting $u_{jk}=-1$
	in a specified row of $z$-bonds and  $u_{jk}=1$ elsewhere. Without loss of generality, we set $u_{jk}=-1$ for $z$-bonds in the $L_y^{th}$ row. Note that we have used the convention that $j$ ($k$) belongs to A (B) sublattice.
	
	\subsection{Itinerant fermions}
	%{\bf{}Itinerant fermions.---}
	As long as the configuration $\{u\}(\Phi_y=-1)$ has been fixed, the effective Hamiltonian for itinerant Majorana fermions ($c^0$) can be rewritten in a standard BdG form,
	\begin{equation}\label{eq:H-Kitaev_BdG}
		\begin{split}
			H_{\text{eff}}\left(\Phi_y=-1\right)=&\sum_{\mathbf{r}}J_x{}\left(2(\eta^0_{\mathbf{r}})^\dagger{}\eta^0_{\mathbf{r}}-1\right)+\sum_{\mathrm{r}_x<L_x,\mathrm{r}_y}J_y{}\left((\eta^0_{\mathbf{r}})^\dagger{}\eta^0_{\mathbf{r}+\hat{\mathbf{x}}}+(\eta^0_{\mathbf{r}})^\dagger{}(\eta^0_{\mathbf{r}+\hat{\mathbf{x}}})^\dagger+\mbox{h.c.}\right)\\
			&+\sum_{\mathbf{r}}\left(J_z(\eta^0_{\mathbf{r}})^\dagger{}\eta^0_{\mathbf{r}+\hat{\mathbf{y}}}+\left(J_z+2i{J_3}\right)(\eta^0_{\mathbf{r}})^\dagger{}(\eta^0_{\mathbf{r}+\hat{\mathbf{y}}})^\dagger+\mbox{h.c.}\right)\left(1-2\delta_{\mathrm{r}_y,L_y}\right),\end{split}
	\end{equation}
	where
	\begin{equation}\label{eq:eta0}
		\eta^0_{\mathbf{r}}=\frac{c^0_{j=(\mathbf{r},\mathrm{A})}+ic^0_{k=(\mathbf{r},\mathrm{B})}}{2}=\frac{i\left(f_{j=(\mathbf{r},\mathrm{A}),\downarrow}^{}-f_{j=(\mathbf{r},\mathrm{A}),\downarrow}^\dagger\right)-\left(f^{}_{k=(\mathbf{r},\mathrm{B}),\downarrow}-f_{k=(\mathbf{r},\mathrm{B}),\downarrow}^\dagger\right)}{2}
	\end{equation}
	is the complex fermion within the unit cell $\mathbf{r}=\mathrm{r}_x\hat{\mathbf{x}}+\mathrm{r}_y\hat{\mathbf{y}}$.
	The above quadratic Hamiltonian can be diagonalized by using standard BdG transformation,
	\begin{equation*}
		H_{\text{eff}}\left(\Phi_y=-1\right) = -\sum_{m=1}^{L_x{}L_y}\epsilon_{m}\left(h^{\dagger}_{m}h_{m}-\frac{1}{2}\right),
	\end{equation*}
	where $\epsilon_{m}\ge{}0$, $h_m^\dagger$ ($h_m$) is a Bogoliubov quasihole creation (annihilation) operator, and its ground state is given by
	\begin{equation}\label{eq:phi0}
		|\phi(\Phi_y=-1)\rangle=\prod_{m=1}^{L_x{}L_y}h^\dagger_m|0\rangle_{\eta^0}
	\end{equation}
	with $|0\rangle_{\eta^0}$ being the vacuum of complex fermions $\eta^0_{\mathbf{r}}$.
	
	\subsection{Gauge fermions}
	%{\bf{}Gauge fermions.---}
	In addition to itinerant complex fermions $\eta^0_{\mathbf{r}}$, one can construct three complex gauge fermions per unit cell, which are defined on every nearest neighboring (NN) bond $\langle{}jk\rangle\in{}a$ ($a=x,y,z$) as
	\begin{equation}\label{eq:eta_a}
		\eta^a_{jk} =\frac{c^a_j-i{}c^a_k}{2},
	\end{equation}
	where $j$ ($k$) belongs to A (B) sublattice.
	It is easy to verify that ${u}_{jk}\equiv{}ic_j^{a}c_k^{a}=1-2(\eta^a_{jk})^\dagger{}\eta^a_{jk}$,
	which indicates that a configuration $\{u\}$ can be obtained in accordance to the occupation number of gauge complex fermions $\eta^a_{jk}$ on every NN bond. Explicitly, an occupied bond $(\eta^a_{jk})^\dagger{}\eta^a_{jk}=1$ gives rise to $u_{jk}=-1$
	, while an empty bond $(\eta^a_{jk})^\dagger{}\eta^a_{jk}=0$ gives rise to $u_{jk}=1$. For instance, the gauge configuration for $\{u\}(\Phi_y=1)$ is obtained by filling all the $z$-bonds in the $L_y$-th row and leaving other bonds empty.

	\begin{figure}
		\includegraphics[scale=0.075]{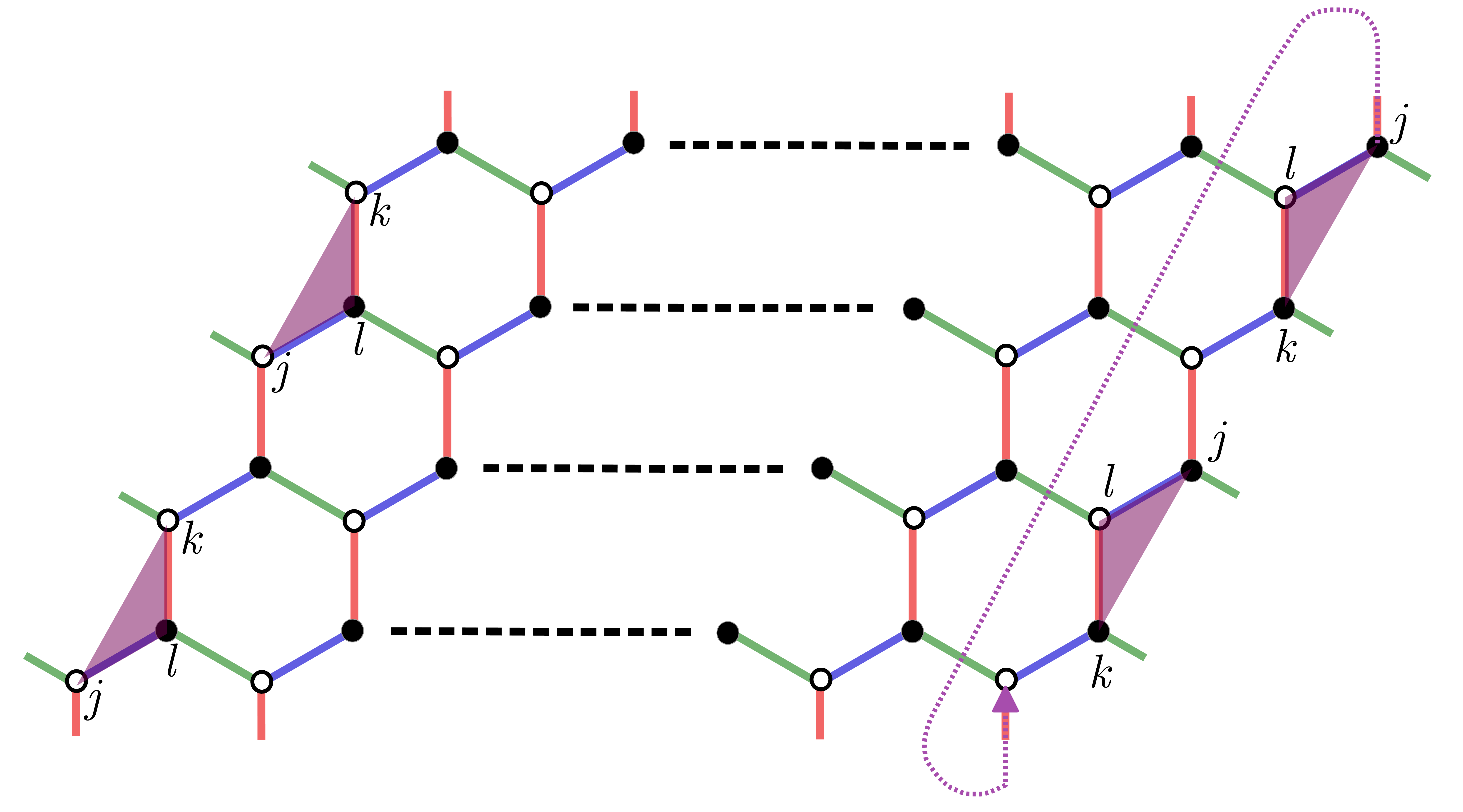}
		\caption{ The three-spin interactions in Eq.~\eqref{eq:H_boundary} are defined on two types of boundary triangles (on the first and last columns, respectively) with vertices $j$, $k$, and $l$.}\label{fig:H_boundary}
	\end{figure}
	As discussed in the main text, the cylindrical boundary condition yields boundary gauge Majorana fermions [see Fig.~2(b) in main text], which should be paired into complex fermions $f_{\langle\langle{}jk\rangle\rangle}$.
	Taking into account all these gauge complex fermions, we obtain the ground state $|u(\Phi_y=-1)\rangle$ as follows,
	\begin{equation}\label{eq:u0}
		|u(\Phi_y=-1)\rangle=\prod_{\langle{}jk\rangle\in{}z,k_y=L_y}(\eta^z_{jk})^\dagger|0\rangle_{\eta},
	\end{equation}
	where $|0\rangle_{\eta}$ is the vacuum of complex fermions $\eta^{a}_{jk}$ and $f_{\langle\langle{}jk\rangle\rangle}$, and $k_y$  is the $y$-component of the unit cell vector on which lattice site $k$  locates.
	Note that we leave all the boundary gauge fermions $f_{\langle\langle{}jk\rangle\rangle}$ unoccupied in Eq.~\eqref{eq:u0} as mentioned in the main text,
	{ and such a state is stabilized by the following boundary Hamiltonian:
		\begin{equation}
			\mathcal{H}_{\text{boundary}}=\delta{}K\sum_{\langle\langle{}jlk\rangle\rangle}\sigma^z_j\sigma^y_l\sigma^x_k=\delta{}K\sum_{\langle\langle{}jlk\rangle\rangle}u_{jl}u_{lk}ic^y_jc^y_k.\label{eq:H_boundary}
		\end{equation}
		Note that $0<\delta{}K\ll1$, $[\mathcal{H}_{\text{boundary}}, \mathcal{H}_3]=0$ , and $\langle\langle{}jlk\rangle\rangle$ refers to three sites around  the boundary triangles as indicated in Fig.~\ref{fig:H_boundary}.
		Indeed there exists other choices, e.g., filling all the the boundary gauge fermions $f_{\langle\langle{}jk\rangle\rangle}$ gives rise to the following state:
		\begin{equation*}
			|u(\Phi_y=-1)\rangle'=\prod_{\langle\langle{}jk\rangle\rangle}f^\dagger_{\langle\langle{}jk\rangle\rangle}\prod_{\langle{}jk\rangle\in{}z,k_y=L_y}(\eta^z_{jk})^\dagger|0\rangle_{\eta},
		\end{equation*}
		which is degenerate with $|u(\Phi_y=-1)\rangle$ if $\delta{}K=0$.}
	
	\subsection{Vacuum states}
	%{\bf{}Vacuum states.---}
	Note that we have chosen the vacuum of itinerant fermions $\eta^0$ in Eq.~\eqref{eq:phi0} and the vacuum of gauge fermions $\eta_{jk}$ and $f_{\langle\langle{}jk\rangle\rangle}$ in Eq.~\eqref{eq:u0}, respectively. However, all the gauge fermions are defined on each bond. To implement the Gutzwiller projection on each site, we had better work on the basis of Abrikosov fermions $f_{j,\uparrow(\downarrow)}$. As shown in Ref.~[\onlinecite{MPOMPS2}], the vacuum state $|0\rangle_{\eta^0}$ could be replaced by the vacuum state of Abrikosov fermions, $|0\rangle$, as long as they have the same fermion parity. Furthermore, the vacuum state  $|0\rangle_{\eta}$ can be obtained by applying annihilation operators $\eta_{jk}$ and $f_{\langle\langle{}jk\rangle\rangle}$ onto the Abrikosov fermion vacuum $|0\rangle$. Thus, we can use the following unprojected ground state in the MPO-MPS calculation:
	\begin{equation}\label{eq:psi0f}
		|\Psi_0(\Phi_y=-1)\rangle=\prod_{m=1}^{L_x{}L_y}h^\dagger_m\prod_{\langle\langle{}jk\rangle\rangle}f_{\langle\langle{}jk\rangle\rangle}\prod_{\langle{}jk\rangle\in{}y,k_x<L_x}\eta^y_{jk}\prod_{\langle{}jk\rangle\in{}x}\eta^x_{jk}\prod_{\langle{}jk\rangle\in{}z,k_y=L_y}(\eta^z_{jk})^\dagger\prod_{\langle{}jk\rangle\in{}z,k_y<L_y}\eta^z_{jk}|0\rangle,
	\end{equation}
	where the operators $h^\dagger_m$, $f_{\langle\langle{}jk\rangle\rangle}$, $\eta^{x,y,z}_{jk}$ and $(\eta^z_{jk})^\dagger$ should be further expressed as linear combinations of Abrikosov fermions $f_{j,\uparrow(\downarrow)}$.

	\subsection{The sequence of acting operators}
	%{\bf{}The sequence of acting operators.---}
	With the help of Eqs.~\eqref{eq:c2a} and \eqref{eq:psi0f}, one can rewrite the Majorana ground state $|\Psi_0\rangle$ as a paired state of Abrikosov fermions, which can be converted  to an MPS by using the MPO-MPS method~\cite{MPOMPS2}.
	We apply the single-particle operators successively in the sequence that is illustrated in Fig.~\ref{fig:sequence}. It turns out that this specified sequence reduces the entanglement in the MPO-MPS process and gives rise to the smallest accumulated truncation error $\epsilon_{\text{trunc}}$.
	
	\begin{figure}[hptb]
		\includegraphics[scale=0.5]{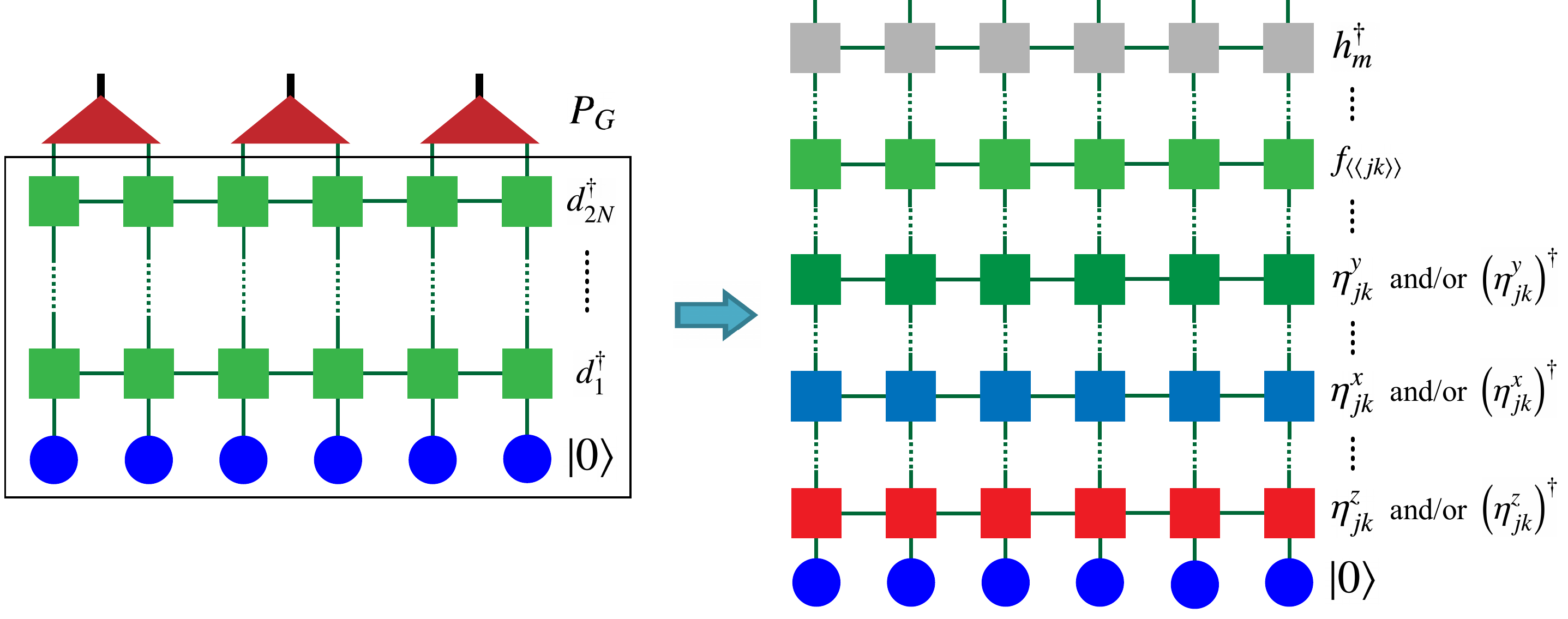}
		\caption{The single-particle operators are acted in the following sequence: gauge complex fermions $\eta^z$ [and/or $\left(\eta^z\right)^\dagger$], $\eta^x$ [and/or $\left(\eta^x\right)^\dagger$], and $\eta^y$ [and/or $\left(\eta^y\right)^\dagger$], boundary complex fermion $f^\dagger$, and finally the Bogoliubov quasiholes $h^\dagger$ (whose sequence is subject to the Wannier localization and ``left-meet-right'' schemes~\cite{MPOMPS1,MPOMPS2}) associated with the effective Hamiltonian $H_{\text{eff}}$.  }\label{fig:sequence}
	\end{figure}
	
	It is also worth noting that, instead of $h^\dagger_m$  themselves, we use linear combinations of of $h^\dagger_m$ to form maximally localized Wannier orbitals~\cite{MPOMPS1,MPOMPS2} and act them on the MPS during the whole MPO-MPS procedure. Additionally, the ``left-meet-right'' scheme~\cite{MPOMPS1,MPOMPS2} has been used.
	Here the leftmost site is labeled by $\tilde{j}=1$ and rightmost site by $\tilde{j}=N=2L_x{}L_y$ [see Eq.~\eqref{eq:label12} for the definition of $\tilde{j}$].
	
	\subsection{Exact zero mode}
	%{\bf{}Exact zero mode.---}
	Consider the sector $\Phi_y=1$. For the non-Abelian phase along the symmetric line $J_x=J_z$, there exists an exact zero mode associated with $c^0$ Majorana fermions on the cylinder geometry. This zero mode gives rise to an additional two-fold degeneracy in the spectrum and has to be properly handled.
	
	The analytical form of the zero mode can be explicitly obtained. For that, we perform Fourier transformation for the Hamiltonian $H_{\text{eff}}(\Phi_y=1)$ along the $y$-direction,
	\begin{equation}
		\begin{split}
			H_{\text{eff}}(\Phi_y=1)=&\sum_{q_y}\sum_{\mathrm{r}_x}J_x{}\left(2(\eta^0_{q_y,\mathrm{r}_x})^\dagger{}\eta^0_{q_y,\mathrm{r}_x}-1\right)+J_y{}\left((\eta^0_{q_y,\mathrm{r}_x})^\dagger{}\eta^0_{q_y,\mathrm{r}_x+1}+(\eta^0_{q_y,\mathrm{r}_x})^\dagger{}(\eta^0_{-q_y,\mathrm{r}_x+1})^\dagger+\mbox{h.c.}\right) \\
			& +J_z\left\{\left[(\eta^0_{q_y,\mathrm{r}_x})^\dagger{}\eta^0_{q_y,\mathrm{r}_x}+\left(1+i\frac{J_3}{J_z}\right)(\eta^0_{q_y,\mathrm{r}_x})^\dagger{}(\eta^0_{-q_y,\mathrm{r}_x})^{\dagger}\right]e^{iq_{y}}+\mbox{h.c.}\right\},
		\end{split}\label{eq:H-Kitaev_BdG_ky}
	\end{equation}
	where $q_y=2n_y\pi/L_y$ for $n_y=1,\dots,L_y$ and $\eta^0_{q_y,\mathrm{r}_x}$ is obtained by taking the Fourier transformation of $\eta^0_{\mathbf{r}}$ [see Eq.~\eqref{eq:eta0}] along the $y$-direction.
	Here we have chosen a gauge such that $u_{jk}=1$ everywhere to stay in the $\Phi_y=1$ sector.
	Because $J_x=J_z$, the $q_y=\pi$ sector of Hamiltonian $H_{\text{eff}}$ in Eq.~\eqref{eq:H-Kitaev_BdG_ky} is equivalent to the Kitaev's Majorana chain, which reads
	\begin{equation}
		h_{\text{eff}}\left(\Phi_y=1,q_y=\pi\right)=\sum_{\mathrm{r}_x=1}^{L_x-1}J_y{}\left((\eta^0_{\pi,\mathrm{r}_x})^\dagger{}\eta^0_{\pi,\mathrm{r}_x+1}+(\eta^0_{\pi,\mathrm{r}_x})^\dagger{}(\eta^0_{\pi,\mathrm{r}_x+1})^\dagger+\mbox{h.c.}\right). \label{eq:Kchian}
	\end{equation}
	There exist two unpaired Majorana fermions
	$\gamma_1=\frac{1}{2}\left((\eta^0_{\pi,1})^\dagger+\eta^0_{\pi,1}\right)$ and $\gamma_{L_x}=\frac{i}{2}\left((\eta^0_{\pi,L_x})^\dagger-\eta^0_{\pi,L_x}\right)$
	commuting with $h_{\text{eff}}$ in Eq.~\eqref{eq:Kchian}.
	Then, pairing up $\gamma_1$ and $\gamma_{L_x}$ gives one complex fermion zero mode $f^\dagger_{1,L_x}=(\gamma_1\pm{}i\gamma_{L_x})$.
	
	To perform the MPO-MPS procedure for the Hamiltonian $H_{\text{eff}}(\Phi_y=1)$ with this exact zero mode, we need to pin the zero mode $f^\dagger_{1,L_x}$ into its vacuum (i.e., annihilated by $f_{1,L_x}$). Otherwise, the MPO-evolved state has an odd fermion parity (measured in terms of the original Abrikosov fermions) and cannot survive the Gutzwiller projection requiring even fermion parity.
	
	\section{Numerical data for Fig.~3 in the main text}
	
	The relative energy deviation $\delta{}E_g$ [defined in Eq.~(6) in the main text] is plotted in Fig.~3 in the main text. Here we list corresponding data in Table~\ref{tab:dEg}.
	The calculations are performed for the Hamiltonian $\mathcal{H}_3$ on an $L_x \times L_y =6 \times 6$ cylinder with parameters $J_x=J_y=J_z=1$ and $J_3=0$. The bond dimension is chosen as $\tilde{D}=200$ for initial MPSs.
	
	The exact ground-state energies are $E_g(\Phi_y=1)=-54.469847490329$ and  $E_g(\Phi_y=-1)=-54.385529432483$.
	Note that $\delta E_g$ initialized with a random MPS is measured from $E_g(\Phi_y=-1)$.

	\begin{table}[htbp]
		\renewcommand\arraystretch{1.55}
		\setlength\tabcolsep{0.15cm}
		\begin{tabular}{c|c|c|c|c|c|c|c}
			\hline
			\hline
			&   \multicolumn{2}{c|}{$P_G|\Psi_0(\Phi_y=1)\rangle$}  &   \multicolumn{2}{c|}{$P_G|\Psi_0(\Phi_y=-1)\rangle$}  &    &\multicolumn{2}{c}{Random MPS} \\
			\hline
			$N_{\text{S}}$ &  $\delta{E}_g$ & $D$ & $\delta{E}_g$ & $D$ & $N_{\text{S}}$ & $\delta{E}_g$  & $D$ \\
			\hline
			7 & 6.3835$\times{}10^{-8}$ & 3875 & 1.1948$\times{}10^{-7}$ & 4216 & 17 & 2.2756$\times{}10^{-4}$ & 4230 \\
			8 & 2.0020$\times{}10^{-8}$ & 4632 & 2.9244$\times{}10^{-8}$ & 5169 & 18 & 2.0524$\times{}10^{-5}$ & 4621 \\
			9 & 1.0119$\times{}10^{-8}$ & 5120 & 1.1545$\times{}10^{-8}$ & 5929 & 19 & 4.3845$\times{}10^{-6}$ & 5439 \\
			10 & 7.1220$\times{}10^{-9}$ & 5345 & 8.4900$\times{}10^{-9}$ & 6328 & 20 & 1.0873$\times{}10^{-6}$ & 5488 \\
			11 & 5.8850$\times{}10^{-9}$ & 5537 & 6.4520$\times{}10^{-9}$ & 6500 & 21 & 3.9420$\times{}10^{-7}$ & 7190 \\
			12 & 5.2980$\times{}10^{-9}$ & 5638 & 6.2710$\times{}10^{-9}$ & 6500 & 22 & 2.3145$\times{}10^{-7}$ & 8000 \\
			13 & 5.2980$\times{}10^{-9}$ & 5699 & 6.2520$\times{}10^{-9}$ & 6500 & 23 & 1.8347$\times{}10^{-7}$ & 8000 \\
			14 & 5.2770$\times{}10^{-9}$ & 5727 & 6.2470$\times{}10^{-9}$ & 6500 & 24 & 1.6716$\times{}10^{-7}$ & 8000 \\
			15 & 5.2670$\times{}10^{-9}$ & 5784 & 6.2460$\times{}10^{-9}$ & 6500 & 25 & 1.5727$\times{}10^{-7}$ & 8000 \\
			16 & 5.2370$\times{}10^{-9}$ & 5846 & 6.2420$\times{}10^{-9}$ & 6500 & 26 & 1.5291$\times{}10^{-7}$ & 8000 \\
			\hline
			\hline
		\end{tabular}
		\caption{{
				The relative energy deviations $\delta{}E_g$ as plotted in Fig. 3 in
				the main text. $N_\text{S}$ is the number of DMRG sweeps, and $D$ is the bond dimension for DMRG calculations.} }\label{tab:dEg}
	\end{table}

	This section is devoted to the four classes of Gutzwiller projected wave functions used in the main text.
	Essentially, all these states are Gutzwiller projected states of Abrikosov fermions. Corresponding unprojected states are either ground states of BdG type Hamiltonians or a Fermi sea of Abrikosov fermions. Below we shall define these unprojected states $|\Psi_0\rangle$ one by one.
	
	{\em (1) Kitaev non-Abelian state with Chern number $C=1$.} This class of states has been discussed in the main text as well as Sec.~\ref{sec:MPOMPS}. The unprojected state is given by a direct product $|\Psi_0\rangle =|\{u\}\rangle\otimes{}|\phi(\{u\})\rangle$, namely, Eq.~(3) in the main text. As discussed in Sec.~\ref{sec:MPOMPS}, $|\{u\}\rangle$ is determined by filling gauge complex fermions in accordance with the eigenvalue of $u_{jk}$ on each bond, and $|\phi(\{u\})\rangle$ is determined by the effective Hamiltonian $H_{\text{eff}}$ in Eq.~(2) in the main text after fixing the eigenvalues $\pm{}1$ for $u_{jk}$. With the help of Eq.~\eqref{eq:c2a}, $|\Psi_0\rangle$ can be written as a paired state of Abrikosov fermions and realized by filling all the Bogoliubov quasiholes as given in Eq.~\eqref{eq:psi0f}.
	
	In practice, we choose parameters $J_x=1.05$, $J_y=J_z=1$ and $J_3=0.05$ in the Hamiltonian $H_{\text{eff}}$ and set $u_{jk}=1$ for all bonds to ensure $\Phi_y=1$. Note that this choice of $J_{x,y,z}$ avoids the zero mode issue mentioned in Sec.~\ref{sec:MPOMPS}, which occurs only along the symmetric line $J_x=J_z$.
	The bond dimension is chosen to be $\tilde{D}=800$ for an $L_x\times{}L_y= 4 \times 10$ cylinder, which gives rise to truncation error $\epsilon_{\text{trunc}}\sim{}10^{-3}$ in the MPO-MPS process.

	{\em (2) Partially polarized state with $C=1$ and (3) Fully polarized state with $C=0$.}
	As proposed in Ref.~[\onlinecite{QHWang2019}], the unprojected state $|\Psi_0\rangle$ for these two classes can be unified as the ground state of the following BdG-type mean-field Hamiltonian:
	\begin{equation}\label{eq:Z2}
		\begin{split}
			H_{\text{BdG}}=&\sum_{\mathbf{r}}
			\alpha^\dagger_{\mathbf{r}}\left[h_\sigma(\sigma^x+\sigma^y+\sigma^z)+h_\tau\sigma^0\right]\alpha_\mathbf{r}+\beta^\dagger_{\mathbf{r}}\left[h_\sigma(\sigma^x+\sigma^y+\sigma^z)+h_\tau\sigma^0\right]\beta_\mathbf{r}+\left[\alpha_\mathbf{r}^\dagger{}(ih_\tau\sigma^y+h_\tau\sigma^y)\beta_\mathbf{r}^\dagger+\text{h.c.}\right]\\
			&+\sum_{\mathbf{r}}\sum_{\delta\mathbf{r}=\hat{\mathbf{0}},\hat{\mathbf{x}},\hat{\mathbf{y}}}
			\left[-\alpha_{\mathbf{r}+\delta\mathbf{r}}^\dagger{}[t_z(\delta\mathbf{r})\sigma^z+t_0\sigma^0]\beta_\mathbf{r}+\alpha_{\mathbf{r}+\delta\mathbf{r}}^\dagger{}[t_x(\delta\mathbf{r})\sigma^z-t_y(\delta\mathbf{r})\sigma^0]\beta_\mathbf{r}^\dagger+\mbox{h.c.}\right],
		\end{split}
	\end{equation}
	where $\alpha_{\mathbf{r}} = \left(f_{\mathbf{r},\mathrm{A},\uparrow}\ f_{\mathbf{r},\mathrm{A},\downarrow}\right)$ and $\beta_{\mathbf{r}}=\left(f_{\mathbf{r},\mathrm{B},\uparrow}\ f_{\mathbf{r},\mathrm{B},\downarrow}\right)$ are two doublets of Abrikosov fermions on A and B sublattices, respectively, $\mathbf{r}$ labels a unit cell, and the parameters are simplified as follows:
	\begin{equation*}
		\begin{array}{cc}
			t_x(\hat{\mathbf{0}})=t_y(\hat{\mathbf{x}})=t_z(\hat{\mathbf{y}})=t_\parallel, & t_x(\hat{\mathbf{x}})= t_x(\hat{\mathbf{y}})=t_x(\hat{\mathbf{0}})= t_x(\hat{\mathbf{y}})=t_x(\hat{\mathbf{0}})= t_x(\hat{\mathbf{x}})=t_\perp.
		\end{array}
	\end{equation*}
	Then, the Gutzwiller projected state $P_G|\Psi_0(h_\sigma,h_\tau,t_0,t_\parallel,t_\perp)\rangle$ serves as a variational wave function for the Kitaev honeycomb model under a [111] magnetic field.
	By varying the five real numbers $\{h_\sigma,h_\tau,t_0,t_\parallel,t_\perp\}$, the energy is optimized to obtain the best approximation to the ground state by using variational Monte Carlo (VMC) method.
	The VMC-optimized parameters have been obtained in Ref.~[\onlinecite{QHWang2019}].
	
	\begin{figure}[thpb]
		\includegraphics[width=17.5cm]{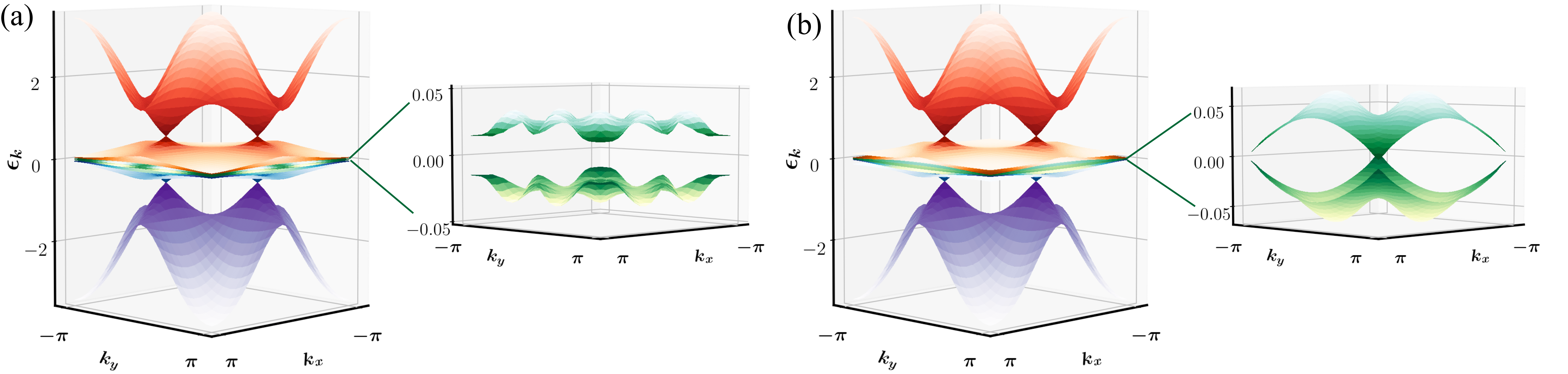}
		\caption{Four quasiparticle bands and four quasihole bands from the Hamiltonian $H_{\text{BdG}}$ defined in Eq.~\eqref{eq:Z2} with (a) parameters given in Eq.~\eqref{eq:c=1} and (b) parameters given in Eq.~\eqref{eq:c=0}. The Chern numbers from the lowest band to the highest band are (a) $\{1,0,0,0,0,0,0,-1\}$ and (b) $\{1,0,0,-1,1,0,0,-1\}$, respectively. Thus, the total Chern number, given by the summation over the Chern numbers of all the quasihole bands, reads (a) $C=1$ and (b) $C=0$.}\label{fig:bands}
	\end{figure}
	
	It is worth noting that, when $h_\sigma=h_\tau=0$ and $t_\perp=-t_0$, the Gutzwiller projected state $P_G|\Psi_0(h_\sigma,h_\tau,t_0,t_\parallel,t_\perp)\rangle$ gives rise to the exact ground state (in $\Phi_y=1$ sector) of the Kitaev honeycomb model $\mathcal{H}_3$ defined in Eq.~(1) in the main text.

	A typical partially polarized state with Chern number $C=1$ [state (2)] is given by the following set of parameters:
	\begin{subequations}
		\begin{equation}\label{eq:c=1}
			\begin{array}{ccccc}
				h_\sigma=-0.341, & h_\tau=0.288, & t_0=-0.588, & t_\parallel=0.622, &t_\perp=0.518,
			\end{array}
		\end{equation}
		while a typical fully polarized state with Chern number $C=0$ [state (3)] is given by another set of parameters:
		\begin{equation}
			\begin{array}{ccccc}\label{eq:c=0}
				h_\sigma=-0.355, & h_\tau=0.276, & t_0=-0.595, & t_\parallel=0.609, &t_\perp=0.524,
			\end{array}
		\end{equation}
	\end{subequations}
	The corresponding quasiparticle and quasihole band structures for Eqs.~\eqref{eq:c=1} and \eqref{eq:c=0} are illustrated in Figs.~\ref{fig:bands}(a) and (b), respectively.
	
	For carrying out the MPO-MPS procedure, the bond dimension $\tilde{D}=1400$ is chosen for an $L_x\times{}L_y= 10 \times 4$ cylinder, which gives rise to truncation errors $\epsilon_{\text{trunc}}\sim{}0.12$ for state (2) and $\epsilon_{\text{trunc}}\sim{}0.09$ for state (3).

	\begin{figure}[htpb]
		\includegraphics[width=16cm]{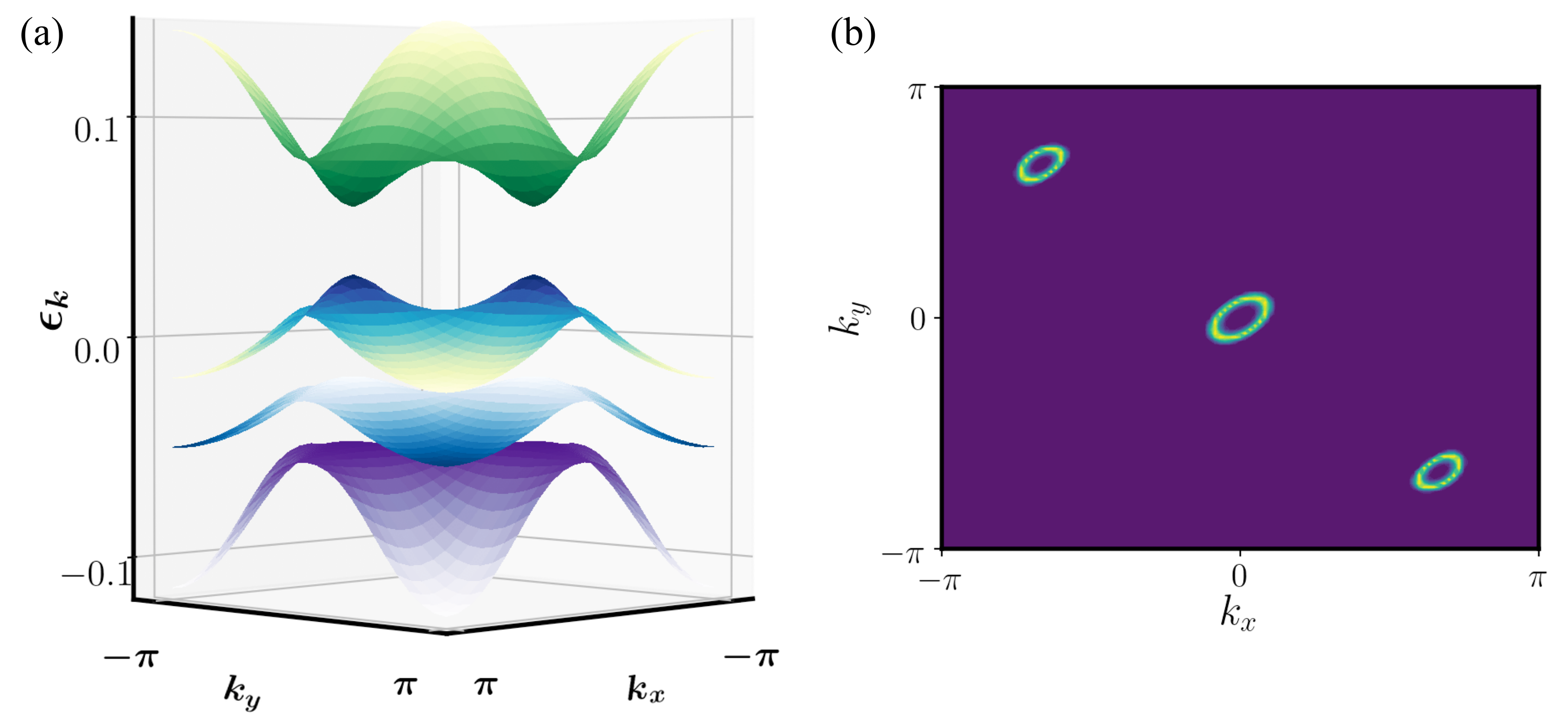}
		\caption{(a) Four spinon bands with a spinon Fermi surface from Hamiltonian $H_{\text{U1}}$ defined in Eq.~\eqref{eq:HU1} with parameters given in Eq.~\eqref{eq:U1param}. (b) The corresponding spinon Fermi surfaces. There are one electron pocket around $(0, 0)$ and two hole pockets around $(\pi,\pi)$.}
		\label{fig:u1band}
	\end{figure}
	
	{\em (4) $U(1)$ spin liquid state with a spinon Fermi surface.} This class of states is introduced in Ref.~[\onlinecite{YMLu2018}] and is labeled as $U1A_{k=0}$ quantum spin liquid state\cite{YMLu2018} in accordance with the projective symmetry group analysis\cite{PSG1,PSG2,PSG3}. The corresponding unprojected ground state $|\Psi_0\rangle$ is given by a quadratic Hamiltonian as follows:
	\begin{equation}\label{eq:HU1}
		\begin{split}
			&H_{U_1}= \sum_{\mathbf{r}}\left\{\alpha_{\mathbf{r}}^\dagger\left[-\frac{h}{8\sqrt{3}}\left(\sigma^x+\sigma^y+\sigma^z\right)-\mu\sigma^0\right]\alpha_{\mathbf{r}}+\beta_{\mathbf{r}}^\dagger\left[-\frac{h}{8\sqrt{3}}\left(\sigma^x+\sigma^y+\sigma^z\right)-\mu\sigma^0\right]\beta_{\mathbf{r}}\right\} \\
			&-\sum_{\mathbf{r}}\left[\alpha^\dagger_{\mathbf{r}}(s_3\sigma^0+t^x_0\sigma^x+t^y_0\sigma^y+t^y_0\sigma^z)\beta_{\mathbf{r}}+\alpha^\dagger_{\mathbf{r}+\hat{\mathbf{x}}}(s_3\sigma^0+t^y_0\sigma^x+t^x_0\sigma^y+t^y_0\sigma^z)\beta_{\mathbf{r}}+\alpha^\dagger_{\mathbf{r}+\hat{\mathbf{y}}}(s_3\sigma^0+t^y_0\sigma^x+t^y_0\sigma^y+t^x_0\sigma^z)\beta_{\mathbf{r}}+\text{h.c.}\right] \\
			&-\sum_{\mathbf{r}}\left\{\left(\alpha^\dagger_{\mathbf{r}+\hat{\mathbf{x}}}, \; \beta^\dagger_{\mathbf{r}+\hat{\mathbf{x}}}\right)\left[(\tilde{s}_3\sigma^0-i\tilde{s}_0\sigma^z)\otimes{}\sigma^0+(\tilde{t}_0^x\sigma^0-i\tilde{t}_3^x\sigma^z)\otimes{}\sigma^x+(\tilde{t}_0^x\sigma^0-i\tilde{t}_3^x\sigma^z)\otimes\sigma^y+(\tilde{t}_0^z\sigma^0-i\tilde{t}_3^z\sigma^z)\otimes\sigma^z\right]\left(\begin{array}{c}
				\alpha_{\mathbf{r}}\\\beta_{\mathbf{r}}
			\end{array}\right)+\mbox{h.c.}\right\} \\
			&-\sum_{\mathbf{r}}\left\{\left(\alpha^\dagger_{\mathbf{r}+\hat{\mathbf{y}}}, \; \beta^\dagger_{\mathbf{r}+\hat{\mathbf{y}}}\right)\left[(\tilde{s}_3\sigma^0-i\tilde{s}_0\sigma^z)\otimes{}\sigma^0+(\tilde{t}_0^x\sigma^0-i\tilde{t}_3^x\sigma^z)\otimes{}\sigma^x+(\tilde{t}_0^z\sigma^0-i\tilde{t}_3^z\sigma^z)\otimes\sigma^y+(\tilde{t}_0^x\sigma^0-i\tilde{t}_3^x\sigma^z)\otimes\sigma^z\right]\left(\begin{array}{c}
				\alpha_{\mathbf{r}}\\\beta_{\mathbf{r}}
			\end{array}\right)+\mbox{h.c.}\right\} \\
			&-\sum_{\mathbf{r}}\left\{\left(\alpha^\dagger_{\mathbf{r}+\hat{\mathbf{x}}-\hat{\mathbf{y}}}, \; \beta^\dagger_{\mathbf{r}+\hat{\mathbf{x}}-\hat{\mathbf{y}}}\right)\left[(\tilde{s}_3\sigma^0-i\tilde{s}_0\sigma^z)\otimes{}\sigma^0+(\tilde{t}_0^z\sigma^0-i\tilde{t}_3^z\sigma^z)\otimes{}\sigma^x+(\tilde{t}_0^x\sigma^0-i\tilde{t}_3^x\sigma^z)\otimes\sigma^y+(\tilde{t}_0^x\sigma^0-i\tilde{t}_3^x\sigma^z)\otimes\sigma^z\right]\left(\begin{array}{c}
				\alpha_{\mathbf{r}}\\\beta_{\mathbf{r}}
			\end{array}\right)+\mbox{h.c.}\right\},
		\end{split}
	\end{equation}
	where real numbers $\{s_3,t^x_0,t^y_0\}$ and $\{\tilde{s}_0,\tilde{s}_3,\tilde{t}_0^x,\tilde{t}_3^x,,\tilde{t}_0^z,\tilde{t}_3^z\}$ are mean-field couplings on nearest neighbor (NN) and next NN bonds, respectively.
	Notice that the geometry of the honeycomb lattice and, thereby, the Hamiltonian $H_{U_1}$ in Eq.~\eqref{eq:HU1} are different from those used in Ref.~[\onlinecite{YMLu2018}] by a global $\pi/3$ rotation.
	
	A typical $U(1)$ spin liquid state with a spinon Fermi surface [state (4)] is obtained by the following set of parameters:
	\begin{equation}
		\begin{split}\label{eq:U1param}
			&h=-0.188,\quad \mu=-0.1832,\quad s_3=-0.02373,\quad t^x_0=-0.02373,\quad  t^y_0=-0.00415,\quad \\
			&\tilde{s}_0=-0.00018,~~~\tilde{s}_3=0.0018,~~~\tilde{t}_0^x=0,~~~\tilde{t}_3^x=-0.000102,~~~\tilde{t}_0^z=0,~~~\tilde{t}_3^z=-0.000102.
		\end{split}
	\end{equation}
	Note that the parameter $h$ in Ref.~[\onlinecite{YMLu2018}] reads 0.188, but the actual value of $h$ which is utilized for practical calculations should be  $-0.188$.
	The corresponding band structures for Eq.~\eqref{eq:U1param} are illustrated in Fig.~\ref{fig:u1band}.
	In the MPO-MPS procedure, the bond dimension is chosen to be $\tilde{D}=800$ for an $L_x\times{}L_y= 10 \times 4$ cylinder, which gives rise to a truncation error $\epsilon_{\text{trunc}}\sim{}0.01$.
	
	{
		\section{$\{W_y,W_p\}$-fixed ansatz}
		This section is devoted to discussing efficiency of initializing DMRG using a $\{W_y,W_p\}$-fixed ansatz. In Ref.~\cite{lee2019}, it was shown that the so-called ``loop gas'' tensor network ansatz is an excellent trial wave function for the Kitaev honeycomb model with $J_x=J_y=J_z$. However, it is computationally expensive to encode such a loop gas ansatz tensor network state on a $L_x \times L_y$ cylinder into an MPS since the bond dimension of MPS is $\tilde{D}=7^{L_y}$. Instead, we can initiate an MPS, namely $\{W_y,W_p\}$-fixed MPS, which is the eigenstate of Wilson loop operators $W_y$ and hexagonal plaquettes $W_p$ with all $1-|\Phi_y| < 10^{-15}$ and $1-|w_p| < 10^{-15}$. Here $[W_p, \mathcal{H}_3]=0$, $W_p^2=1$, and $W_p\equiv{}\sigma^{x}_{p_1}\sigma^{y}_{p_2}\sigma^{z}_{p_3}\sigma^{z}_{p_4}\sigma^{y}_{p_5}\sigma^{z}_{p_6}$, where  the site indices $p_1–p_6$ are as $2,3,4,10,11,12$ in Fig.~\ref{fig:quasi1d} (b). We denote $w_p$ as the eigenvalue of $W_p$ and, for the ground states of $\mathcal{H}_3$, all of $w_p$ are 1.
		\begin{figure}[tb]
			\includegraphics[width=10cm]{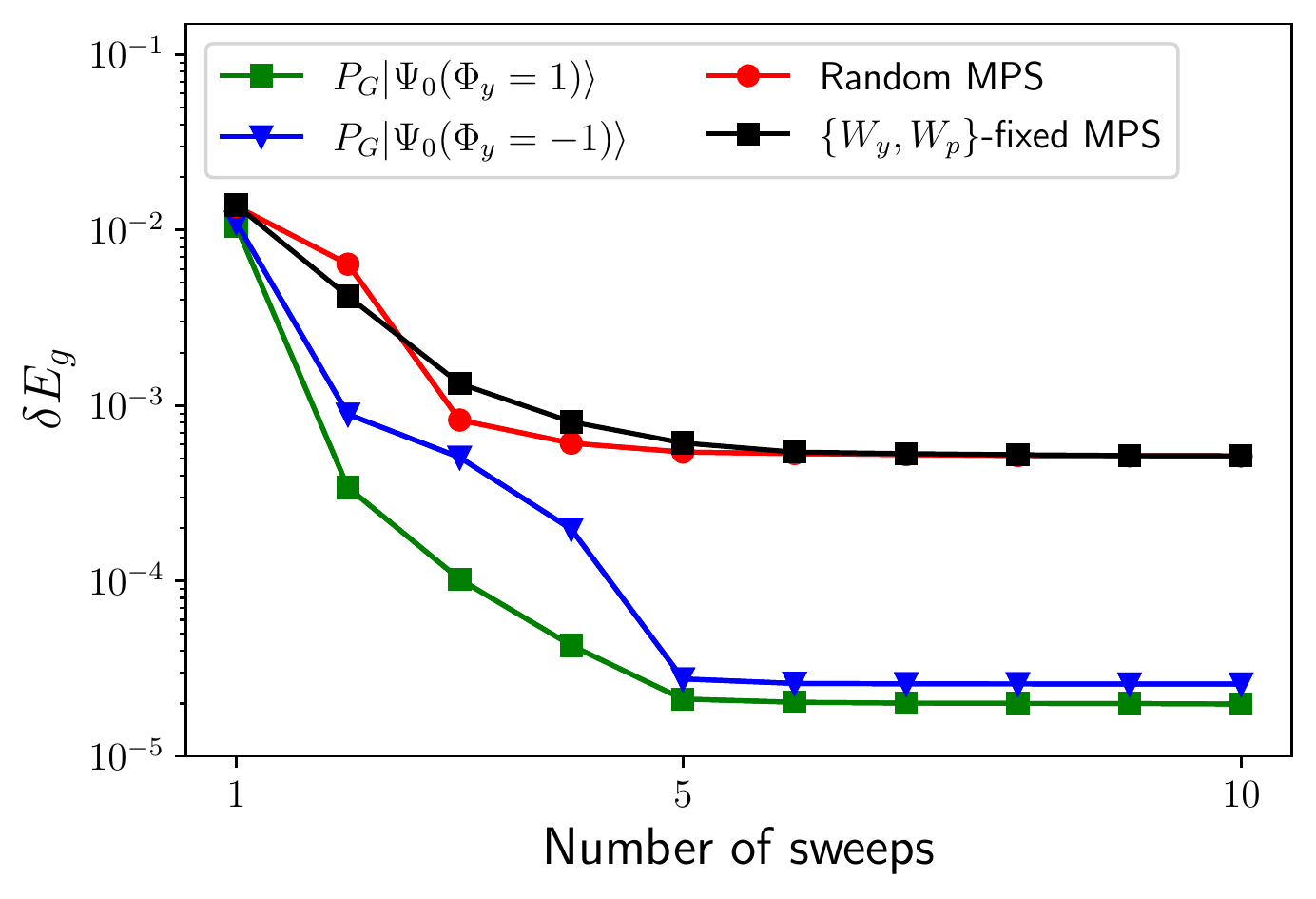}
			\caption{The relative energy deviations $\delta{}E_g$ [defined in Eq.~\eqref{eq:dEg}] versus number of sweeps in DMRG. The calculations are performed for the Hamiltonian $\mathcal{H}_3$ in Eq.~\eqref{eq:H-Kitaev} on an $L_x \times L_y =6 \times 6$ cylinder and with parameters $J_x=J_y=J_z=1$ and $J_3=0$.
				Red, green, blue, and black lines stand for those with initial states of random MPS, $P_G|\Psi_0(\Phi_y=-1)\rangle$, $P_G|\Psi_0(\Phi_y=1)\rangle$, and $\{W_y, W_p\}$-fixed MPS, respectively.
				The bond dimension is chosen as $\tilde{D}=200$ for initial MPSs, and the first 5 DMRG sweeps are used to  gradually increase the bond dimension from $\tilde{D}$ to $D$.
				Note that $\delta E_g$ initialized with a random MPS and a $\{W_y, W_p\}$-fixed MPS are measured from the ground-state energy in the $\Phi_y=-1$ sector. The final bond dimension after DMRG sweeps is $1000$.}
			\label{fig:dEg_loopgas}
		\end{figure}
		
		As illustrated in Fig.~\ref{fig:dEg_loopgas}, the relative energy deviation $\delta{}E_g$ of the $\{W_y,W_p\}$-fixed MPS with $D=1000$ is almost the same as that of the random MPS, which indicates that this initial state is not as efficient as Gutzwiller projected ansatz.
		Moreover, for this $\{W_y,W_p\}$-fixed MPS, it is possible that flipping fluxes $w_p$ does not cost energy due to that itinerant Majorana degrees of freedom are not fixed (and hence the vison gap is zero). Consequently, the Wilson loop $\Phi_y$ would not be preserved during DMRG sweeps. Actually, we have encountered such situations in our numerical simulations.
		
		However, the loop-gas ansatz in Ref.~\cite{lee2019} is not supposed to suffer from this issue, since it is expected to capture the essential entanglement structure of the actual ground state. It would be interesting if one could find a reliable MPS approximation of this ansatz and test its performance in initializing DMRG calculations.
		
		\section{entanglement spectra for Kitaev's non-Abelian states}
		\begin{figure}
			\includegraphics[width=10cm]{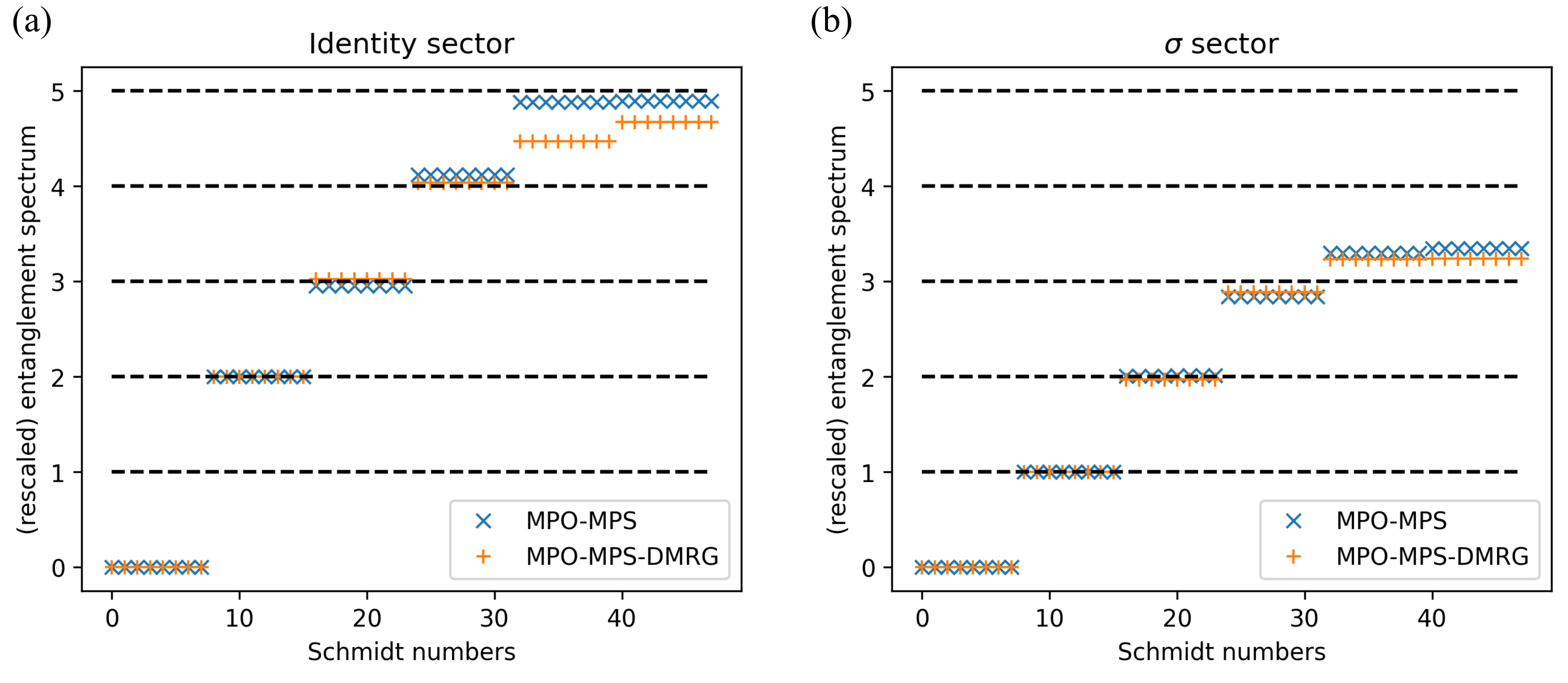}
			\caption{The rescaled entanglement spectra for Gutzwiller wave functions $P_G|\Psi_{0}(\Phi_{y}=-1)\rangle$ ($\mathbbm{1}$ sector) and $P_G|\Psi_{0}(\Phi_{y}=1)\rangle$ ($\sigma$ sector), as well as their corresponding DMRG-optimized states. The calculations are performed on an $L_x\times{}L_y=6\times{}6$ cylinder.}
			\label{fig:ees}
		\end{figure}
		
		In this section, we discuss the topological properties of non-Abelian states where $c^0$ Majoranas have a nontrivial topological band structure with Chern number $C=1$.
		In the Kitaev's $B$ phase~\cite{Kitaev06}, where finite $J_3$ opens a bulk gap, $P_G|\Psi_{0}(\Phi_{y}=-1)\rangle$ and $P_G|\Psi_{0}(\Phi_{y}=1)\rangle$ are denoted by the topological quasiparticles $\mathbbm{1}$ and $\sigma$~\cite{tu2013b}, respectively.
		The rescaled entanglement spectra for $\mathbbm{1}$ and $\sigma$ have been calculated for both initial Gutzwiller projected state (MPO-MPS) at $\tilde{D}=2000$ and DMRG-optimized state (MPO-MPS-DMRG) at $D=2000$, as shown in Fig.~\ref{fig:ees}.
		For both $\mathbbm{1}$ and $\sigma$ sectors, the characteristic counting of entanglement spectra agrees with the prediction of Ising conformal field theory (up to trivial multiplicity, which arises due to the entanglement cut of $\mathbbm{Z}_2$ gauge fields).
		This means that the topological order is well captured by the MPO-MPS method.
		%This characteristic counting still exist after DMRG optimization (see Fig.~\ref{fig:ees}), which indicates that the topological orders will be preserved during DMRG sweeps.
		The similarity of entanglement spectra resulting from the MPO-MPS and MPO-MPS-DMRG calculations indicates that the topological sector is preserved during the DMRG optimization procedure.
	}
	
\end{widetext}

\end{document}